\documentclass{article}
\usepackage{PRIMEarxiv}
\usepackage[T1]{fontenc}    
\usepackage{hyperref}       
\usepackage{url}            
\usepackage{booktabs}       
\usepackage{amsfonts}       
\usepackage{nicefrac}       
\usepackage{microtype}      
 \usepackage{graphicx} 
 \graphicspath{{images/}} 
\usepackage{lipsum}
\usepackage{amsmath}
\usepackage{listings}
\usepackage{xltabular, booktabs}
\usepackage{xcolor}
\lstdefinelanguage{Solidity}{
    keywords={function, public, view, returns, require, emit, uint256, bool, address, mapping},
    keywordstyle=\color{blue}\bfseries,
    ndkeywords={block, timestamp, msg, sender},
    ndkeywordstyle=\color{purple}\bfseries,
    identifierstyle=\color{black},
    sensitive=false,
    comment=[l]{//},
    morecomment=[s]{/*}{*/},
    commentstyle=\color{green}\ttfamily,
    stringstyle=\color{red}\ttfamily,
    morestring=[b]',
    morestring=[b]"
}
\definecolor{keywordblue}{RGB}{0,0,255}
\definecolor{commentgreen}{RGB}{0,128,0}
\definecolor{stringred}{RGB}{163,21,21}
\definecolor{background}{RGB}{245,245,245}

\lstdefinelanguage{JavaScript}{
    keywords={let, await, async, for, while, if, try, catch, throw, else, console, function, return},
    keywordstyle=\color{keywordblue}\bfseries,
    ndkeywords={deployed, getAccounts, isSuccess, mineBlock, getMinerBlockNumbers, getTotalBlockCount, toString},
    ndkeywordstyle=\color{keywordblue}\bfseries,
    commentstyle=\color{commentgreen}\ttfamily,
    stringstyle=\color{stringred}\ttfamily,
    numbers=left,
    numberstyle=\tiny\color{gray},
    frame=single,
    breaklines=true,
    showstringspaces=false
}

\usepackage{booktabs}
\usepackage{enumitem} 
\usepackage{lastpage}   
\usepackage{fancyhdr}       
\usepackage{graphicx}       
\graphicspath{{media/}}     
\usepackage{booktabs} 
\usepackage{longtable}
\usepackage{indentfirst} 
\usepackage{indentfirst} 
\usepackage{tabularx} 
\graphicspath{{images/}}
\usepackage{titlesec}
\titleformat{\subsubsection}
  {\normalfont\bfseries\itshape} 
  {\thesubsubsection}            
  {1em}                          
  {}                             

\titlespacing{\subsubsection}
  {0pt}     
  {1ex}     
  {0.5ex}   
\usepackage{float}
\usepackage{listings}
\usepackage{xcolor}
\fancypagestyle{plain}{%
  \fancyhf{\headrule} 
\fancyfoot[LO]{Wenwen Zhou et.al.: \itshape Preprint Submitted to Arxiv.\\ \textsuperscript{*}These authors contributed equally to this work. }
\fancyfoot[Ro]{Page \thepage\ of \pageref{LastPage}} 
}

\title{BLOCKCHAIN SECURITY BASED ON CRYPTOGRAPHY: A REVIEW}

\author{
  Wenwen Zhou\textsuperscript{*}, \\
  School of Cyberspace Security\\
  Hainan University \\
  Haikou\\
  \texttt{1792301715@qq.com} \\
   \And
     Dongyang Lyu\textsuperscript{*}, \\
  School of Cyberspace Security\\
  Hainan University \\
  Haikou\\
  \texttt{dongyanglyu@hainanu.edu.cn} \\
  \And
  Xiaoqi Li \\
 School of Cyberspace Security \\
  Hainan University \\
 Haikou\\
  \texttt{csxqli@ieee.org} \\
}

\begin{document}
\maketitle 
\thispagestyle{plain}
\begin{abstract}
As an emerging service framework built by combining cryptography, P2P network, consensus mechanism and innovative contract technology, blockchain has been widely used in digital finance, data sharing, message traceability and electronic evidence preservation because of its decentralised, non-tamperable and transaction traceability. However, with the complex and changeable application scenarios of blockchain technology and the continuous enhancement of blockchain attack technology, the security of the blockchain system has been seriously threatened, dramatically affecting the development and application of blockchain technology.
This paper aims to analyse the attacks on blockchain from the perspective of cryptography. Firstly, from the cryptography technology in the blockchain, the principle of hash functions, digital signatures, and other technologies, as well as their role in the blockchain, are introduced. Then, based on the six-layer architecture of the blockchain, the attacks on the data layer, the network layer, the consensus layer, the contract layer, the incentive layer and the application layer are analysed, and the methods to mitigate or resist the attacks are proposed. Secondly, the attack principles of 51\% attack, double-spend attack, re-entrancy attack, replay attack, Sybil attack and timestamp tampering attack were analysed, and the mitigation or defence solutions for these six attacks were designed. Finally, the core problems to be solved in blockchain technology are summarised, and the future development of blockchain security technology is projected.
\end{abstract}


\section{Introduction}
Blockchain technology originated from Satoshi Nakamoto's paper on Bitcoin published in 2008\cite{nakamoto2008bitcoin}, which constructs a secure and trustworthy distributed computing framework through the comprehensive use of cryptography, consensus algorithms, distributed networks, and smart contracts. The core characteristics of blockchain include decentralization, de-trusting, non-tampering, traceability, and anonymity, which makes blockchain provide innovative solutions in solving the problems of deposit, trust, and data governance in the economy and society, and is now widely used in many fields such as digital finance, digital government, Internet of Things (IoT), intelligent manufacturing, supply chain management, etc., which demonstrates its significant value in modern society \cite{zhu2020research}.

However, with the wide application of blockchain technology, its security problems are increasing, especially in the protection of data privacy, identity authentication and transaction security face many threats, the security incidents occurring in various blockchain systems are increasing year by year, the representative major security incidents in the history of blockchain development\cite{shen2024survey} are shown in Table \ref{tab:1}(The Renminbi(RMB) is the official name of China's currency. ). As the foundation of the data economy, the security of the blockchain system has a significant impact on the financial industry, and enhancing the security of the blockchain has become a crucial direction for its development. The blockchain system primarily relies on cryptographic technology to ensure the security and integrity of data. Therefore, studying the principles of cryptography technology and its applications in blockchain has both theoretical and practical significance. Moreover, an in-depth study of the security threats faced by blockchain and the proposal of solutions to resist attacks can effectively enhance the security of blockchain systems, thus promoting the healthy development of blockchain technology.

\begin{table}[ht]
  \centering  
  \caption{Security Event in the History of Blockchain Development}  
  \label{tab:1}  
\begin{tabularx}{\textwidth}{l X X }
 \toprule
  \textbf{Time} & \textbf{Event} & \textbf{Influence}\\
  \midrule
  December 2013 & At least five user's accounts on the Okcoin exchange were stolen at the same time & A total of nearly 640,000 RMB worth of Bitcoin and Litecoin were los  \\
  February 2014 &The world's largest Bitcoin trading platform, Mt.Gox, was attacked by hackers & 850,000 bitcoins were stolen, with a loss worth nearly 500 million US dollars \\
  January 2015 & TBitstamp, the world's third-largest Bitcoin exchange, was hacked & Lost 19,000 bitcoins\\
  June 2016 & The Ethereum star project The DAO was hacked & About 3.6 million Ether were stolen  \\
  July 2017 & Parity, a multi-signature wallet service provider, was attacked by hackers &153,000 Ether was stolen \\
  January 2018 &The Japanese Bitcoin exchange Coincheck was hacked & 523 million NEM coins were stolen, resulting in a loss worth 530 million US dollars \\
  November 2019 &The South Korean cryptocurrency exchange UpBit was hacked & 34,000 Ether were stolen \\
  December 2020 & The veteran public chain Aeternity was attacked by hackers with a 51\% attack & Lost over 39 million AE tokens \\
  August 2021 & The cross-chain interoperability protocol Poly Network was hacked & Crypto assets worth 610 million US dollars were stolen \\
  March 2022 &The blockchain project platform Ronin was attacked by hackers &173,000 Ether and 25.5 million USDC were stolen, with a loss value of approximately 625 million US dollars\\
  March 2023 & The DeFi lending protocol Euler Finance suffered a flash loan attack & Millions of dollars worth of DAI, USDC, StETH and WBTC were stolen, causing losses of approximately 197 million US dollars \\
  May 2024 &The long-established Japanese cryptocurrency exchange DMM Bitcoin was hit by a hacker attack & 4,02.9 bitcoins were stolen, resulting in a loss of over 300 million US dollars \\
  February 2025 &The Security multi-signature wallet on the cryptocurrency exchange Bybit was hacked & 499,000 Ether were stolen, resulting in a loss of approximately 1.5 billion US dollars  \\
  \bottomrule
\end{tabularx}
\end{table}

Tan Minsheng et al\cite{tan2024improvement} proposed an improved Proof-of-Work (POW)-based blockchain consensus mechanism
, called IPOW, which can effectively reduce the possibility of malicious nodes, and reduce the difference in bookkeeping
rights caused by the lack of arithmetic power, and improve the nodes' participation motivation, and the security of the system
Jiang Yibin et al \cite{jiang2024efficient} proposed a Delegated Proof-of-Statement (DPoS) consensus mechanism based on a multi-community chain, called SMDPoS, which outperforms the traditional DPoS mechanism in terms of transaction processing capability and fault handling, proposed a multi-community chain-based delegated proof of ingestion (DPoS) consensus mechanism called SMDPoS, which is superior to the traditional DPoS mechanism in terms of transaction processing capacity and fault tolerance, and can effectively reduce the risk of malicious nodes. Xie Hao-meng \cite{xie2023research} , proposed SofitMix, a mixed-coin protocol that supports off-chain parallel payments, which adopts a hash time-lock and Digital signature technique. The protocol adopts hash time lock and digital signature technology to ensure the atomicity of transactions, and zero-knowledge proof technology to ensure the unlinkability of traders, which can effectively resist DoS attacks and collusion attacks. Hao-Meng Xie also proposes the SPCEX cross-chain protocol in the field of cross-chain technology \cite{xie2024spcex}, which integrates mixed-coin and homomorphic re-encryption technology to protect the privacy of users and transactions, and can effectively resist collusion against pre-run attacks and Sybil attacks, ensuring the security of cross-chain transactions. It can effectively resist collusion, front-running attacks, and Sybil attacks, ensuring cross-chain transaction security.

In recent years, the problem of "fake nodes" has become increasingly prominent. G.M. et al\cite{m2025efficient} achieved intelligent identification of Sybil nodes through BEP20 wallet address generator, Edwards-curve digital signature algorithm, and monitor table, effectively defending Sybil attack. Yang Xinle et al \cite{yang2019effective} proposed a PoW consensus mechanism that introduces history-weighted information to reduce the risk of a 51\%  attack effectively. Somdip Dey \cite{dey2018securing} effectively prevented the 51\% attack in the coalitional blockchain network by integrating machine learning and algorithmic game theory. Haibo Yi et al \cite{yi2021efficient} developed a threshold signature scheme based on the NP complex problem. They utilised the signatures generated by this algorithm in the consensus mechanism, thus enhancing the blockchain system's resistance to quantum attacks.Y. Lu et al \cite{lu2024tree} proposed a tree-based oblivious random access protocol, Tree-ORAP, which implements an access protection model in the privacy-preserving blockchain system. Yi Qian et al \cite{dai2022ddos} created a DDoS attack traffic detection strategy using a cross multilayer convolutional neural network CMCNN model in the network layer of blockchain architecture.

We conducted a more comprehensive and in-depth study on blockchain security based on the above research. Firstly, by collecting actual attack events in blockchain networks in recent years, we summarised the attack methods or security threats used by attackers in each event, and further extracted the 16 most common and representative patterns among them. Meanwhile, based on the recognized level of blockchain architecture, we use Visual Studio Code software to write, compile, deploy, and execute smart contracts in the Windows system environment, combined with Truffle and Ganache, and design a defense against attacks such as the 51\% attack, double-spending attack, reentrancy attack, replay attack, Sybil attack, and timestamp tampering Solution.

\begin{enumerate}[label=\textbullet]
\item[] The main contributions of this paper are as follows:
\item We present an innovative contract solution to mitigate and defend against 51\% attacks, double-spending attacks, reentrancy attacks, replay attacks, Sybil attacks, and timestamp tampering attacks.
\item We verify the executability of contracts in Visual Studio Code software within two different environments (Truffle and Ganache) on the Windows system.

\item We evaluate these contracts in terms of transaction speed, defence capability, resource consumption, and user usability, pointing out their advantages and disadvantages.

\end{enumerate}

\section{BACKGROUND}
\subsection{Related Knowledge}
\subsubsection{Blockchain Cryptography Technology}
A hash function can compress a message of arbitrary length into a fixed-length output value \cite{ahmad2023study} within a limited and reasonable time. The output value is also known as a hash value, and its main form is \textit{h = H(M)},
where \textit{h} is a fixed-length hash value and \textit{M} is a variable-length message. Furthermore, hash functions have the characteristics of one-wayness, collision resistance, determinacy, and avalanche effect \cite{raikwar2019sok}.

\begin{figure}[ht]
  \centering
\includegraphics[width=0.8\textwidth]{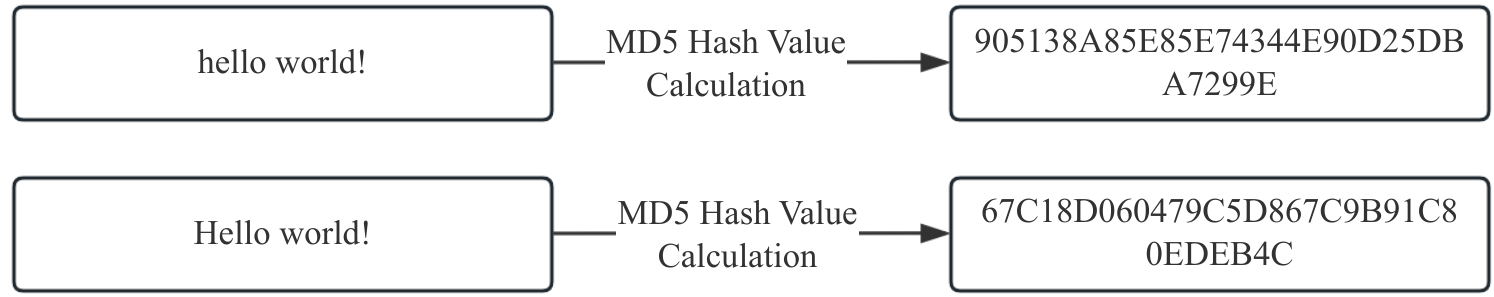}
  \caption{Avalanche Effect of Hash Functions}  
   \label{fig:1}
\end{figure}

The one-way property of a hash function means that for a given $y$, finding an $x$ that satisfies \textit{H($x$) = $y$} is
computationally infeasible. Therefore, calculating the hash value of data is very easy, but it is almost
impossible to reverse engineer the data from the hash value. This phenomenon is called a hash collision when multiple original datasets with differences generate the same
hash value after the same hash operation. Collision resistance is the core security of a hash function. 
All indicators can be divided into two types based on the strength of their defence capabilities: weak collision resistance and strong collision resistance.
The former refers to the computational infeasibility of finding another $x'$ such that \textit{H($x$) = H($x'$)} given $x$.
The latter refers to the computational infeasibility of finding distinct $x$ and $x'$ such that \textit{H($x$) = H($x'$).}
Because hash functions have collision resistance, they prevent attackers from forging.
Data acquisition yields the same hash value. The deterministic characteristic of a hash function means that when the input data is the same,
its output hash value must be consistent. This ensures that the data mapping process is repeatable and
reliable. The avalanche effect refers to the phenomenon where even very slight
differences between the input message value and the target message value will cause the output hash value to deviate significantly and unpredictably from the target hash value.
This high sensitivity makes it difficult for attackers to reverse engineer data from hash values, as shown in Figure \ref {fig:1}.
At the same time, the hash algorithm also has the characteristic of fast verification, which can quickly verify the integrity of the data.
These characteristics make the hash function the basis for blockchain to ensure data integrity and immutability.

The widely used hash functions in cryptography mainly include the MD and SHA series.
SHA-256 is the primary hash algorithm used in blockchain technology systems based on cryptography. It can transform data of any length into a fixed 256-bit output. It is mainly applied to hash links in block headers, Merkle trees \cite{saini2022blockchain}, verifying transaction data's integrity, uniqueness, and immutability.

\subsubsection{Digital Signature}
\begin{figure}[ht]
  \centering
\includegraphics[width=1.0\textwidth]{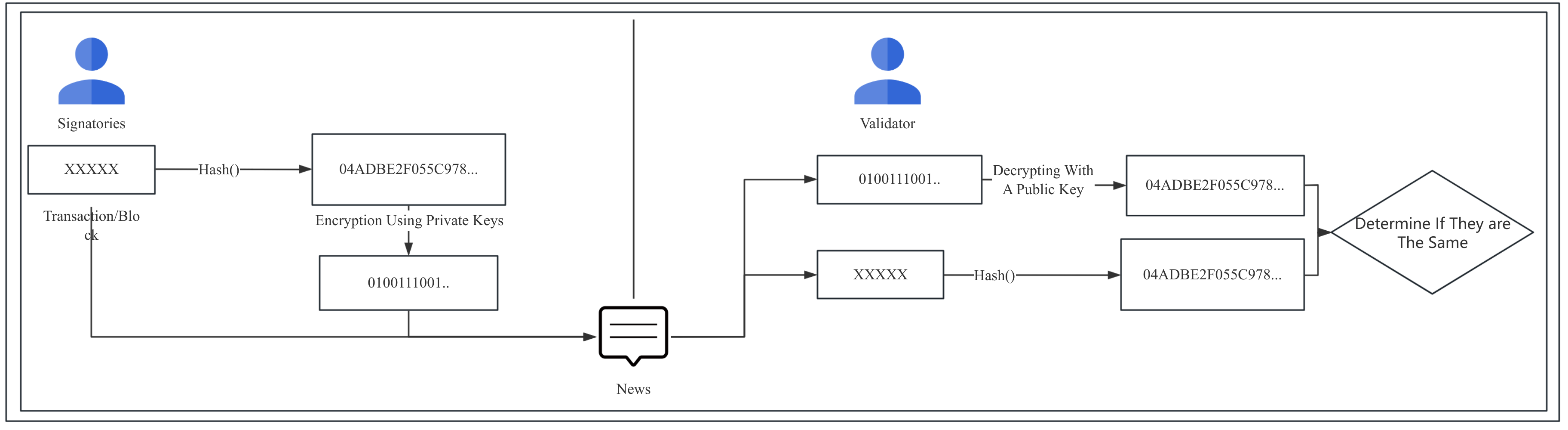 }
  \caption{Digital Signature Signing and Verification Process}  
   \label{fig:2}
\end{figure}

Digital signatures \cite{diffie1976new} are implemented through asymmetric encryption technology and are primarily used to verify the legitimacy of transactions and data integrity. There are three steps to a digital signature: key generation, signing, and verification \cite{cai2021principles}. Figure \ref{fig:2} illustrates the specific signing and verification process.
As the core of the digital signature mechanism, public and private keys are generated strictly according to specific key algorithms. The private key is sensitive information that needs to be stored confidentially by the user to prevent signature invalidation due to private key leakage. In contrast, as public information, the public key must be made public to ensure that any node can verify the digital signature. Moreover, the one-way design of the key algorithm ensures that even if the public key is entirely public, attackers cannot reverse engineer the corresponding private key information, using mathematical principles to ensure the absolute security of the private key.

\begin{figure}[ht]
  \centering
\includegraphics[width=1.0\textwidth]{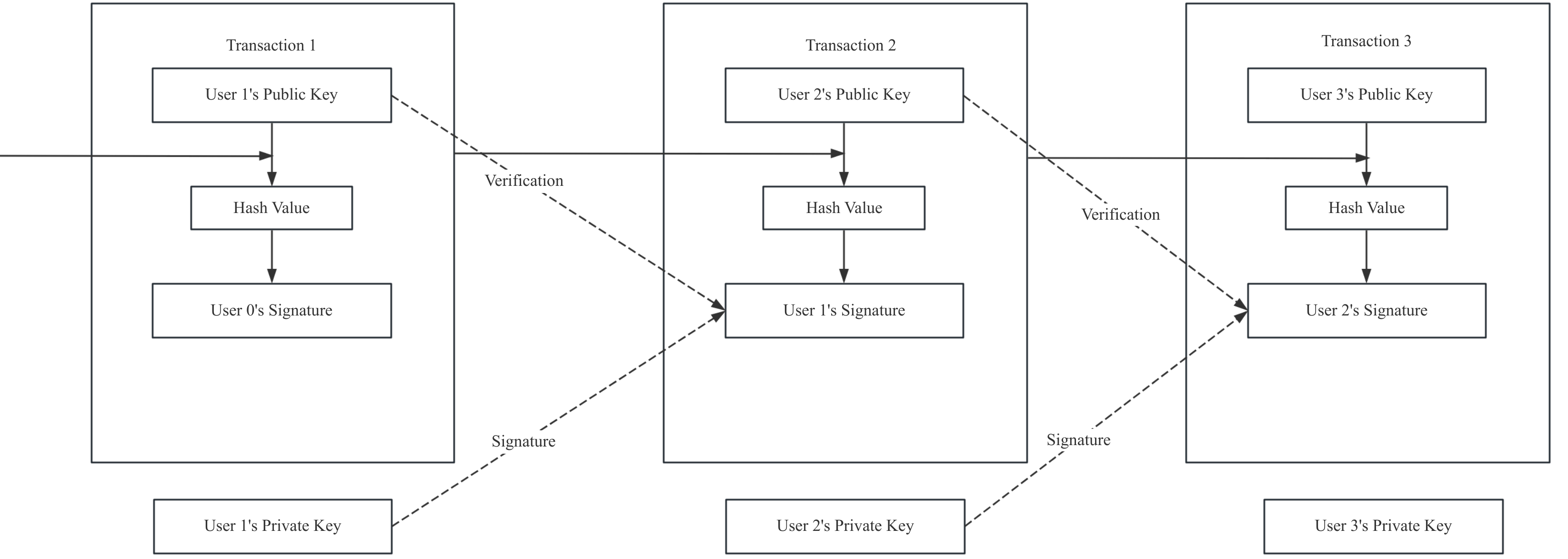}
  \caption{Use of Digital Signatures in Blockchain}  
   \label{fig:3}
\end{figure}

During the signing and verification process, the signer first hashes the message to be signed, generating a fixed-length digest through a one-way function. Then, the signer uses its private key to encrypt the digest, producing a signature credential with authentication capabilities. The message and credentials are then encapsulated and transmitted over the network. After receiving the complete data packet via network transmission, the recipient decrypts the signature using the corresponding public key and restores the initial hash value. Subsequently, the recipient uses the same hash algorithm as the signer to recalculate the hash value of the original message and generate a local digest. Finally, the recipient compares the consistency between the original hash value and the local digest to complete the verification. Suppose the comparison result shows a complete match. In that case, it can be confirmed that the original message has not been maliciously tampered with by attackers during network transmission and that the sender is legitimate. When the digital signature mechanism is applied to the blockchain, each transaction can be traced back to a specific user, thereby realizing the identity authentication of transaction users, the integrity of transaction content, and the non-repudiation of transaction behavior. The use of digital signatures in the blockchain is shown in Figure \ref{fig:3}. 
At the same time, digital signatures can also be used to authenticate node identities and prevent Double-spending attacks.

The digital signature algorithm commonly used in blockchain is the Elliptic Curve Digital Signature Algorithm $ECDSA$\cite{lamba2013efficient}, which
The algorithm performs operations in an elliptic curve over a finite field, assuming the private key of the algorithm is $k$ and the public key is $K$,
$G$ is the base point of the elliptic curve $E_p(a, b)$, $n$ is the order of $G$, and it satisfies $K=k\cdot G$, based on this mathematical structure
The private key signing process is as follows:
 \begin{itemize}
     \item[(1).]Choose a random number $r (r < n)$, and calculate the point $(x, y)$ = $r \cdot G$ and $d$ = $x \bmod n$.
     \item[(2).]Calculate $s$=$r^{-1} \cdot$ $(h+d \cdot k) \bmod n$ based on the random number $r$, private key $k$ and the hash value $h$ of the message $M$.
     \item[(3).]Send the message $M$ and signature $\{d, s\}$ to the receiver.
 \end{itemize}
 After the recipient receives the data, they will perform signature verification. The process of using the public key to verify the signature is as follows:
\begin{itemize}
     \item[(1).]Check if $d$ and $s$ are less than $n$, and calculate the hash value $h$ based on the received message $M$.
     \item[(2).]Calculate $u$=$s^{-1}$ $\bmod$ $n$, then calculate using the sender's public key $K$ to calculate ($x_1$ , $y_1$ ) = $u\cdot h\cdot G+ d\cdot u\cdot K$.
     \item[(3).]Calculate $v$=$x_1$ $\bmod$ $n$ and determine if its value is consistent with $d$. If they are consistent, then the sign Verification is successful, the message has not been tampered with.
 \end{itemize}
 
  \begin{table}[ht]
  \centering  
  \caption{Comparison of Key Lengths for RSA and Elliptic Curve Algorithms with Equivalent Security Strength}  
  \label{tab:2}  
 \begin{tabularx}{\textwidth}{ccc}
 \toprule
  \textbf{ Security Strength(Bit)} & \textbf{RSA Algorithm Key Length(Bit)} & \textbf{Elliptic Curve Cryptography(ECC) Key Length
  (Bits)}\\
  \midrule
  56 & 512 &112 \\
  80 & 1024 &160 \\
  112 & 2048 &224 \\
  128 & 3072 &256 \\
  192 & 7680 &384 \\
  256 & 15360 &512 \\
  \bottomrule
\end{tabularx}
\end{table}
 
The elliptic curve algorithm has significant advantages over the traditional RSA algorithm. While ensuring the same cryptographic strength, this algorithm significantly reduces the required key length for encryption and greatly improves computational efficiency. It features strong encryption and high computational efficiency. Table \ref{tab:2} compares key lengths for RSA and elliptic curve algorithms at the same security strength. In addition to the elliptic curve digital signature algorithm, the digital signature technologies used in blockchain also include Ring Signatures \cite{rivest2006how}, Multi-Signatures \cite{itakura1983public}, Blind Signatures \cite{chaum1983blind}, and Threshold Signatures \cite{desmedt1992shared}. Table \ref{tab:3} compares these four types of signatures.

\begin{xltabular}{\textwidth}{X X X X} 
  \caption{Comparison of Four Signature Technologies} 
  \label{tab:3} \\ 
  \toprule 
  \textbf{Signature Algorithm} & \textbf{Use Cases} & \textbf{Advantages} & \textbf{Disadvantages} \\
  \midrule
  \endfirsthead 
  
  \multicolumn{4}{c}{\textmd{Table \thetable~ (Continued): Comparison of Four Signature Technologies}} \\
  \toprule
  \textbf{Signature Algorithm} & \textbf{Use Cases} & \textbf{Advantages} & \textbf{Disadvantages} \\
  \midrule
  \endhead 
  
  \bottomrule
  \endfoot 
  
  Ring Signature & Anonymous Transactions are achieved without revealing the real signer & Participants can conduct transactions without revealing their identities, protecting their privacy & Allowing anyone in the ring to sign poses a risk of malicious use and is difficult to monitor and investigate \\
  Multi Signature & Multiple private keys are required to sign the transaction & Increase transaction security, decentralize trust and control, and support flexible multi-signature control schemes & Increase transaction time and complexity, and ensure that the private keys of multiple parties are fully protected and managed \\
  Blind Signature & Sign transactions without exposing user data & Protecting user data privacy while ensuring transaction reliability & The inability to confirm the origin and authenticity of the signature poses a challenge to the blockchain verification mechanism \\
  Threshold Signature & Creating a signature using multiple private keys & While improving security and preventing single point failures, it also better protects user privacy & Requires more computing resources and time to create signatures, increasing management difficulty \\

\end{xltabular}

\subsubsection{Data Encryption}
In cryptography, encryption algorithms are divided into two major systems: symmetric and asymmetric. Symmetric encryption technology uses a single key to complete the data encryption and decryption process, which has the characteristics of fast operation speed and is suitable for large-scale data processing. Standard algorithms include AES, DES, and 3DES. However, the symmetric encryption key must be pre-distributed to both communicating parties, which can very easily cause key leakage risks in an open network environment and become a security risk. Asymmetric encryption algorithms, also known as public key encryption, use the public key for encryption and the private key for decryption \cite{zhu2016security}. Although it does not require pre-sharing of keys, its computational efficiency is low. Standard algorithms include RSA and elliptic curve cryptography. Symmetric encryption mechanisms are often used in blockchain technology architectures to protect sensitive data confidentiality. In contrast, asymmetric encryption systems implement security functions such as identity authentication and digital signatures.

While traditional encryption techniques effectively protect data privacy when encrypting data, they also impact data availability. Unlike traditional encryption techniques, homomorphic encryption is a special type that supports direct computation on ciphertext. The decrypted result of the computed ciphertext is consistent with the corresponding computation on the plaintext \cite{liu2024progress}. In distributed application scenarios based on blockchain technology, homomorphic encryption can effectively address the privacy protection of sensitive data and transaction data during the execution of smart contracts, allowing users to use blockchain systems without worrying about privacy breaches.

\subsubsection{Zero-Knowledge Protocol}
\begin{figure}[ht]
  \centering
\includegraphics[width=0.8\textwidth]{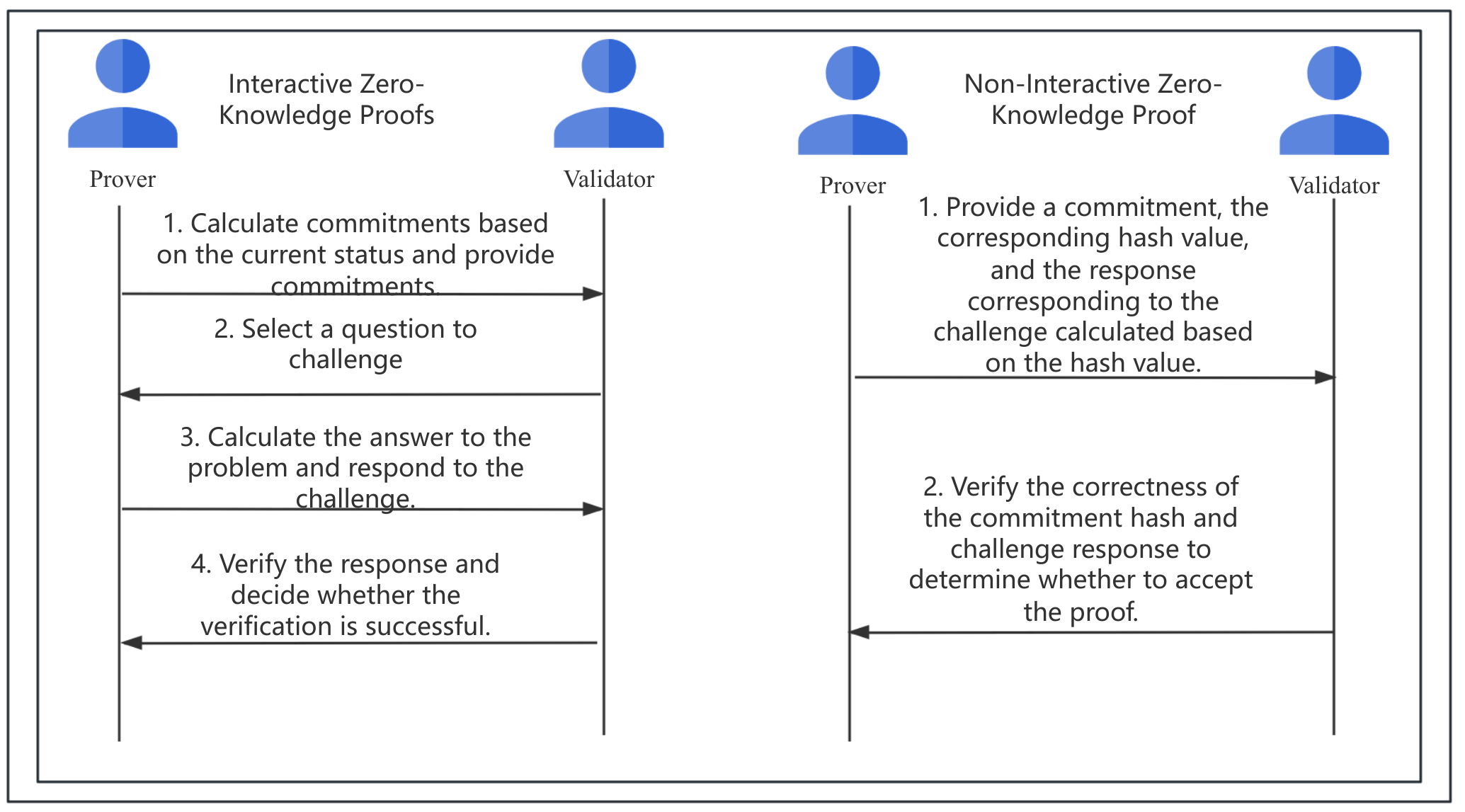}
  \caption{Interactive Zero-Knowledge Proof and Non-Interactive Zero-Knowledge Proof Processes}  
   \label{fig:4}
\end{figure}

Zero-knowledge proof \cite{sasson2014zerocash}, as an important technology in modern cryptography, can prove the authenticity of certain information without revealing specific data. Its core cryptographic principle lies in the prover's ability to verify to the verifier that they possess that specific information without exposing the specific content of the private information they hold. Based on the differences in interaction mode, the protocol can be subdivided into interactive and non-interactive implementation frameworks, with the two types of systems shown in Figure \ref{fig:4}. The former type requires the prover and the verifier to exchange information in multiple rounds to achieve secure verification. In contrast, the latter only requires one information transfer to achieve verification, which is simpler and more efficient, so blockchain typically uses non-interactive zero-knowledge proofs.

Zero-Knowledge Proof technology is crucial to protecting privacy in the architecture of blockchain technology. It is mainly used to keep users' private information confidential or to hide sensitive information while executing smart contracts. Zero-knowledge proof technology can improve the security and privacy of blockchain systems\cite{ben2018scalable}. Common zero-knowledge proof technology protocols include zk-SNARKs\cite{luong2023privacy}, zk-STARKs, BulletProof\cite{bunz2018bulletproofs}, and Sonic\cite{maller2019sonic}. The comparison of these four protocols is shown in Table \ref{tab:4}. ZK-SNARKs in non-interactive Zero-Knowledge Proof Protocols have now become the mainstream cryptographic tool in blockchain systems. By generating only a concise proof, the verifier can determine the proof's authenticity without multiple rounds of information exchange between the two parties. Therefore, this technology is widely used in Ethereum's privacy transactions and smart contract privacy protection. Compared to ZK-SNARKs, ZK-STARKs have stronger resistance to quantum attacks, improving the efficiency and transparency of the proof process.

\begin{xltabular}{\textwidth}{X X X X X} 
  \caption{A Comparison of Four Common Zero-Knowledge Proof Protocol Techniques} 
  \label{tab:4} \\ 
  \toprule 
  \textbf{Scheme} & \textbf{Proof Generation Complexity} & \textbf{Proof Verification Complexity}& \textbf{Proof Size}& \textbf{Trusted Settings}\\
  \midrule
  \endfirsthead 
  
  \multicolumn{5}{c}{\textmd{Table \thetable~ (Continued): A Comparison of Four Common Zero-Knowledge Proof Protocol Techniques}} \\
  \toprule
  \textbf{Scheme} & \textbf{Proof Generation Complexity} & \textbf{Proof Verification Complexity}& \textbf{Proof Size}& \textbf{Trusted Settings}\\
  \midrule
  \endhead 
  
  \bottomrule
  \endfoot 
  
  ZK-SNARKs & O($n$log$N$) &O(1)&O(1)&Yes \\
  ZK-SNARKs & O($n$ ploylog$N$) &O(ploylog$N$)&O(ploylog$N$)&No \\
  BulletProof & O($n$log$N$) &O($N$)&O(log$N$)&No \\
  Sonic & O($n$log$N$) &O(1)&O(1)&Yes \\

\end{xltabular}

\subsubsection{Post-Quantum Cryptography}
Breakthroughs in the field of quantum computing pose a significant challenge to traditional cryptographic systems, threatening the security of blockchain systems. Two algorithms in quantum computing can threaten blockchain networks \cite{yu2024research}: Shor algorithm \cite{shor1994algorithms} and Grover algorithm \cite{grover1996fast}, which can crack traditional encryption algorithms such as RSA and ECC in a short period. Table \ref{tab:5} shows the impact of Shor algorithm on traditional encryption algorithms in quantum computing. The security foundation of contemporary cryptography is mainly built on the computational complexity of specific mathematical problems. These problems constitute the core theoretical support for traditional encryption algorithms, while the Shor algorithm can efficiently crack the mathematical problems of discrete logarithms and large integer factorisation. Therefore, the current mainstream cryptographic algorithms lose their security guarantees in the face of quantum attacks. At the same time, Grover's algorithm reduces the time complexity of finding hash collisions through quantum acceleration, significantly weakening the collision resistance of hash functions. Therefore, under the threat of quantum computing, blockchain systems that rely on public key encryption and hash algorithms will face serious security challenges.

In response to the threat of quantum algorithms to traditional cryptographic systems, the cryptography community has proposed the concept of post-quantum cryptography. This field focuses on developing quantum-Security algorithms to create secure mechanisms capable of withstanding quantum attacks, providing theoretical support and technical securityguards for information security. Post-quantum cryptography encompasses several branches, including lattice-based cryptography, code-based cryptography, multivariate cryptography, hash-based cryptography, and supersingular isogeny elliptic-curve cryptography \cite{alghamdi2021future}. Among these, lattice-based cryptography is currently the most promising research direction in post-quantum cryptography. The post-quantum algorithm standards released by the National Institute of Standards and Technology (NIST) include $CRYSTALS-Kyber$, CRYSTALS-Dilithium\cite{bai2020crystals}, FALCON\cite{fouque2020falcon}, and SPHINCS+\cite{bernstein2019sphincs}, marking substantial progress in post-quantum cryptography. The applications of post-quantum cryptography in blockchain include transaction signatures, data encryption, and identity verification, providing security guarantees against quantum attacks for blockchain \cite{fernandez2020towards}.

\begin{xltabular}{\textwidth}{X X X X} 
  \caption{The Impact of Shor Algorithm on Traditional Encryption Algorithms} 
  \label{tab:5} \\ 
  \toprule 
  \textbf{Scheme} & \textbf{Proof Generation Complexity} & \textbf{Proof Verification Complexity}& \textbf{Proof Size}\\
  \midrule
  \endfirsthead 
  
  \multicolumn{4}{c}{\textmd{Table \thetable~ (Continued): The Impact of Shor Algorithm on Traditional Encryption Algorithms}} \\
  \toprule
\textbf{Scheme} & \textbf{Proof Generation Complexity} & \textbf{Proof Verification Complexity}& \textbf{Proof Size}\\
  \midrule
  \endhead 
  
  \bottomrule
  \endfoot 
  
  Secure Hash Algorithm 256-Bit(SHA-256) & Hash Algorithm &Digital Fingerprint&Security Halved\\
  Advanced Encryption Standard (AES) & Symmetric-Key Algorithm &Encryption&Security Halved \\
 Rivest Shamir Adleman (RSA) & Asymmetric Cryptographic Algorithm Based On Integer Factorization &Encryption, Signature&Complete Failure \\
Digital Signature Algorithm(DSA) & Asymmetric Cryptographic Algorithm Based On Finite Field &Encryption, Signature&Complete Failure \\
Elliptic Curve Digital Signature Algorithm(ECDSA) & Asymmetric Cryptographic Algorithm Based On Elliptic Curve &Encryption, Signature&Complete Failure\\ Elliptic Curve Diffie-Hellman(ECDH)&Asymmetric Cryptographic Algorithm Based On Elliptic Curve&Encryption, Signature&Complete Failure\\

\end{xltabular}

\subsubsection{Public Key Infrastructure}
Public key infrastructure (PKI) is crucial for identity authentication, data encryption, and signature verification in blockchain systems\cite{liu2024progress}. PKI manages digital certificates and keys, which enable data encryption and signature verification. Identity authentication prevents malicious nodes from joining the network and ensures the integrity and authenticity of communication data in blockchain systems.

\begin{figure}[ht]
  \centering
\includegraphics[width=0.8\textwidth]{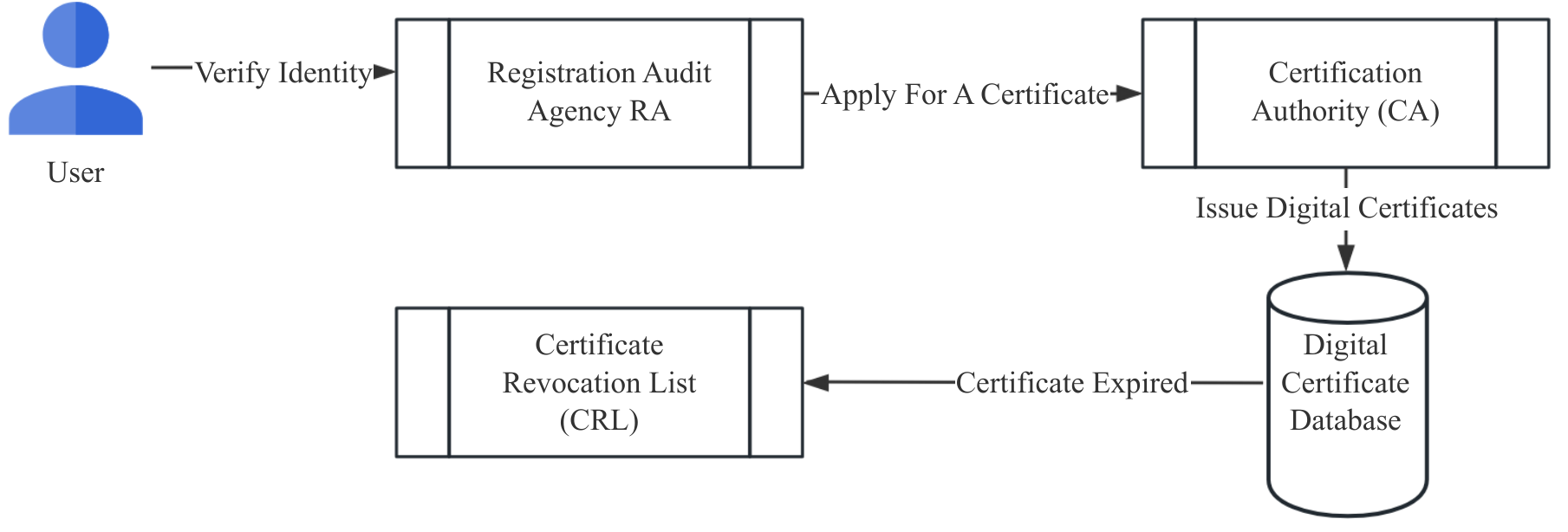}
  \caption{PKI System Operation Process}  
   \label{fig:5}
\end{figure}

The Public Key Infrastructure (PKI) architecture comprises key entities, including the Certificate Authority (CA), Registration Authority (RA), digital certificate database, and Certificate Revocation List (CRL) \cite{zhang2025pki}. Among them, the CA, as a trusted third-party institution, undertakes the critical function of issuing digital certificates and authenticating them. After the Registration Authority (RA) performs multi-dimensional verification of the user's identity and completes the identity verification, it submits the request to the Certificate Authority (CA) for certificate issuance. The digital certificate database stores the issued certificates and manages the certificate lifecycle. If a certificate becomes invalid or is tampered with, the certificate will be revoked through the Certificate Revocation List (CRL). Figure \ref {fig:5} shows the entire PKI system process. The PKI system enables decentralised trust in blockchain systems, providing users with a secure and reliable transaction environment.

\subsection{Related Attacks}
Cryptography is central to blockchain's defence system: zero-knowledge proofs (such as ZK-SNARKs) can verify the authenticity of transactions without revealing data, resisting privacy analysis. homomorphic encryption ensures the privacy of on-chain data computation. and post-quantum cryptography algorithms (such as lattice-based cryptography) are a key path to addressing future threats. However, technology must be combined with auditing (such as formal verification of smart contracts), governance (node admission mechanisms), and user education to build a defence-in-depth system. This article mainly focuses on the following 16 types of attacks, which are divided into four types, as detailed in Table \ref {tab:6}.

\begin{table}[H]
  \centering  
  \caption{Multi-Layer Impact Analysis of Blockchain Vulnerabilities}  
  \label{tab:6}  
 \begin{tabularx}{\textwidth}{X X X X}
 \toprule
  \textbf{SchemeVulnerability Type} & \textbf{Typical Attack Methods} & \textbf{Technical Principles}& \textbf{Impact Dimensions}\\
  \hline
  Basics Of Cryptography & Quantum Computing, Hash Collisions &Algorithm Cracking&Collapse Of Underlying Trust System\\
 Network And Consensus & 51\% Attack&Compute Power Monopoly&Consensus Failure, Double-Spending Risk
 \\
 Smart Contract & Reentrancy, Integer Overflow &Logic Flaws, Numerical Boundary Neglect&Asset Theft, Contract Paralysis\\
Applications And Humans& Social Engineering, Privilege Escalation &Psychological Deception, Misconfiguration&Systemic Data Breach\\
  \hline
\end{tabularx}
\end{table}
\section{Security Analysis}
\begin{figure}[ht]
  \centering
\includegraphics[width=0.8\textwidth]{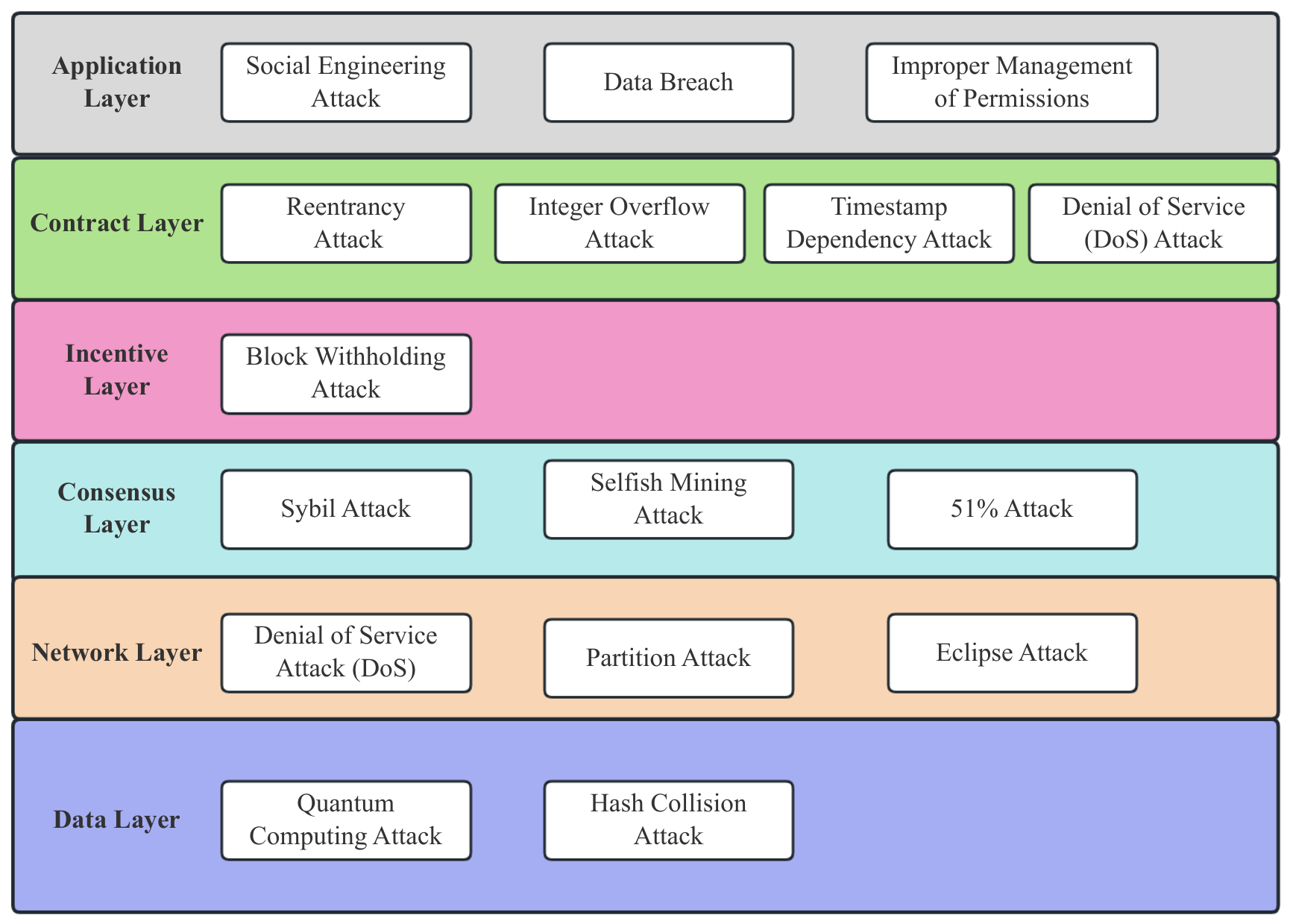}
  \caption{ Main Attacks Suffered by Each Layer of the Blockchain}  
   \label{fig:7}
\end{figure}

  This chapter primarily analyzes the primary security threats and corresponding countermeasures introduced by various layers within the blockchain architecture. Figure \ref{fig:7} illustrates the main attacks suffered by each layer in the six-layer architecture. The basic data layer mainly faces challenges such as quantum computing attacks and hash collision attacks. The network transmission layer needs to guard against denial-of-service attacks, fragmentation attacks, and Eclipse attacks. During operation, the consensus mechanism must resist various security threats, including 51\% attacks, Sybil attacks, and precomputation attacks. The incentive layer faces the risk of block withholding attacks, where miners may maliciously withhold blocks to gain an advantage. The smart contract layer needs to be wary of code defect vulnerabilities, such as reentrancy attacks, timestamp manipulation, DoS attacks, and integer overflow attacks. The application layer faces risks, including phishing attacks, data breaches, and improper permission management. Therefore, by analyzing the attacks originating from different layers and breaking down each layer, we propose implementing distinct protective measures at each level to effectively defend against various potential attack threats, enhance the security of the blockchain system, and ensure the long-term stable and secure operation of the entire blockchain system.

\subsection{Data Layer Threat}
\begin{figure}[ht]
  \centering
\includegraphics[width=1.0\textwidth]{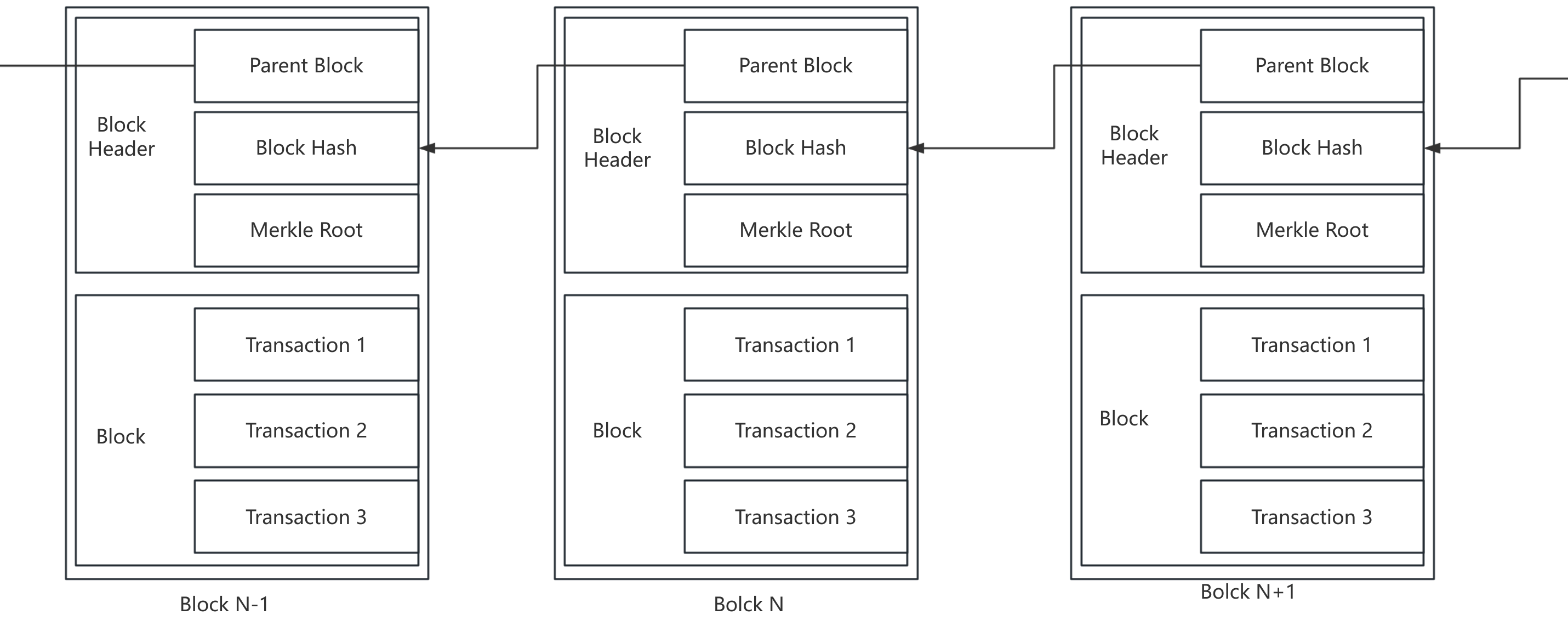}
  \caption{Chain Structure in Blockchain}  
   \label{fig:6}
\end{figure}

The data layer, a fundamental module in blockchain system architecture, constructs the unique data storage structure of blockchain by adopting key technologies such as chain-like data organization, hash algorithms, and Merkle trees. It is primarily responsible for the structured storage of a set of transaction records that have been consensus-validated. Each block is connected through hash pointers, effectively ensuring the integrity of transaction data, the traceability of historical states, and the resistance to tampering of stored information. This provides a solid foundation for the stable operation and secure transactions of the blockchain system. The chain-like structure of blockchain is shown in Figure \ref{fig:6}. However, due to the chain structure of the data layer, which relies on hash value linking, it has security issues, such as quantum computing attacks and hash collision attacks\cite{shen2024survey}.

Grover's algorithm can significantly improve the efficiency of hash operations through quantum acceleration. However, it also exponentially increases the probability of hash collisions, enabling attackers to reconstruct the blockchain through precomputation attacks. Under the guise of maintaining data integrity, attackers can exploit the acceleration properties of quantum computing in the search space to traverse hash structures, thereby breaking existing tamper-proof mechanisms. They can then precisely construct data blocks with the same hash value as the original blocks, thereby hiding tampered transaction histories and undermining the blockchain's immutability. Post-quantumPost-quantum cryptographic algorithms can be introduced to counter quantum computing attacks to counter quantum computing attacks, such as the CRYSTALS-Kyber algorithm for data encryption and the Falcon or $SPHINCS+$ algorithms for digital signatures.

Hash collision attacks, also called hash collision attacks, exploit weaknesses in hash algorithms, allowing multiple different inputs to map to the same hash value. When many colliding hash values are stored in a storage structure, it causes chain accumulation in the data, degrading the originally highly efficient $O(1)$ key-value lookup mechanism into a performance-degraded linear linked list structure. This increases the complexity of the time during operations such as searching and inserting. When the system frequently processes requests, it consumes many system resources, causing the system's response speed to drop sharply and preventing it from handling regular user requests promptly. Eventually, this leads to a denial-of-service (DoS) attack, causing the system to become paralysed. To effectively resist hash collision attacks, the current standard approach is a composite defence strategy. On the one hand, it uses new-generation hash functions with stronger collision resistance. On the other hand, it employs optimised hash table collision resolution mechanisms, such as open addressing, chaining, and dynamic quadratic probing, which disperse collision data when collisions are detected, thereby maintaining the stability of the hash table.

\subsection{Network Layer Threats}
The network layer serves as the core for data interaction and consensus in blockchain technology, leveraging a peer-to-peer (P2P) architecture to achieve decentralisation and ensure equal participation among nodes\cite{wang2023defense}. However, the openness of P2P exposes it to various security threats: Denial-of-Service (DoS) attacks \cite{elleithy2005denial} block transactions by depleting resources, which can be mitigated through traffic monitoring/proof-of-work. partition attacks \cite{tian2021research} disrupt consensus by splitting the network, leading to forks, which can be prevented through multi-path routing/dynamic node adjustments. Inherent $P2P$ flaws (such as lack of authentication and monitoring) allow malicious nodes to freely enter and exit the network, spreading harmful data or abnormal traffic. Security can be enhanced through identity authentication and anomaly detection systems. Eclipse attacks \cite{bhumichai2023feature}control neighbouring nodes of target nodes to receive false data or isolate them, which can be mitigated using diversified connection pools and reputation assessment mechanisms.
\subsection{Consensus Layer Threat}
It primarily faces the threat of Sybil attacks \cite{douceur2002sybil}, where attackers forge multiple node identities to dominate the consensus process and manipulate outcomes (tampering with or reversing transactions). Defence measures include establishing strict node identity authentication (requiring certification by most trusted nodes) and increasing the cost of creating multiple identities. Different consensus mechanisms face specific threats: Proof-of-Work (PoW) mechanisms are susceptible to 51\% attacks \cite{anita2019blockchain} (where controlling over 50\% of computational power allows for chain reorganization to implement double-spending or transaction rejection) and selfish mining attacks \cite{nicolas2021blockchain} (where hiding block construction to build a longer private chain renders legitimate miners' work). The latter can be mitigated through fork detection mechanisms to limit depth defence. Proof-of-Stake (PoS) mechanisms mitigate the resource consumption of PoW but are susceptible to Pre-computation attacks (where attackers exploit token holdings and the hash correlation between blocks to predict and monopolise block production rights). Defense measures include using Verifiable Delay Functions (VDF) to increase prediction difficulty, implementing dynamic stake dilution to reduce the weight of malicious nodes, deploying forward-secure hash chains to weaken hash correlations, or introducing salt values to enhance randomness. The Delegated Proof of Stake (DPOS) mechanism carries the risk of witness nodes being attacked or colluding maliciously, leading to validation failures or data tampering. while the Byzantine Fault Tolerance (PBFT) mechanism offers better security in consortium blockchains, malicious control or manipulation of nodes could still disrupt consensus formation \cite{alghamdi2024survey}.
\subsection{Threats in the Incentive Layer}
The incentive layer of the blockchain, as the driving force of the system operation, will reward the nodes in the network that participate in transaction verification and block generation through economic incentives, which aims to mobilize the motivation of the participants and encourage the nodes to maintain the security of the system, however, the malicious nodes may use some improper means to make economic benefits.

A block deduction attack \cite{bag2017bitcoin} refers to the malicious behavior of an attacker who mines a new block and then intentionally conceals the hash value to construct a longer chain, thereby obtaining higher rewards. This behavior reduces the revenue of the entire mining pool and undermines the fairness of the revenue of other honest miners. In order to effectively prevent block deduction attacks, we can set up a real-time monitoring system for arithmetic and workload to identify abnormal arithmetic fluctuations and abnormal workload submission behavior, and set up a disciplinary incentive mechanism to permanently remove malicious miners detected launching an attack from the network or disqualify their block qualification, to enhance the security and fairness of the system.
\subsection{Contract Layer Threats}
The smart contract layer is the core component of the blockchain system that executes automated contracts, and its tamper-proof nature makes contract vulnerabilities potentially serious. Reentrancy attacks are the most common threat, in which an attacker repeatedly withdraws funds by recursively calling a contract function before updating its state. Timestamp-dependent contracts are susceptible to manipulation, and it is recommended that a trusted third-party time source be introduced to replace blockchain timestamps. Denial-of-service attacks exhaust Gas resources through malicious transactions. Countermeasures include establishing a multi-administrator co-regulation mechanism and implementing multi-signature privilege control. Preventing these security threats requires constructing a defence system at multiple levels, including code auditing, operation monitoring, and governance mechanisms. Secure coding practices at the development stage are especially critical, including using verified security libraries and implementing strict input validation checks. At the same time, establishing a contract upgrade mechanism and an emergency suspension function can also prevent losses promptly when vulnerabilities are exposed. With the continuous expansion of imaginative contract application scenarios, its security will directly affect the healthy development of the blockchain ecosystem, requiring developers, auditing organisations, and users to participate in security governance jointly.

\subsection{Application Layer Threat}
The blockchain application layer serves as a key entry point for user interaction, and its security directly affects user experience and asset security. Social engineering attacks are the most common form of threat, in which attackers induce users to disclose sensitive information such as private keys through phishing websites and forged emails. In addition, front-end interface hijacking and code injection attacks are also common, with attackers manipulating user transactions by tampering with page content or implanting malicious code. In response to these threats, users need to raise their security awareness, cautiously verify transaction information and avoid clicking on suspicious links. The risk of data leakage should also not be ignored. Unencrypted or weakly encrypted transmission data may be intercepted by network sniffing and other means, so it is important to use strong encryption algorithms to protect private data. Vulnerabilities in privilege management may lead to unauthorized operations, requiring the establishment of strict authentication mechanisms and fine-grained privilege control systems. To address these challenges, application layer security requires multi-level measures: including the implementation of end-to-end encrypted transmission, regular security audits, the use of multi-signature mechanisms, as well as the establishment of user security education system. It is worth noting that application layer security requires the joint participation of developers, operators and users to build an all-round security protection system by combining technical protection with security awareness enhancement. With the expansion of blockchain application scenarios, application layer security will become a key link to ensure the healthy development of the entire ecosystem.
\section{Design and Implementation}
\subsection{51\% Attack}
\subsubsection{Principle of Attack}
A 51\% attack in the blockchain is a scenario where an individual or group controls more than 50\% of the computing power or mining hash rate, enabling them to manipulate the transaction network and cause damage to the network.
Such as undoing transactions, double-spending attacks, and preventing other nodes from adding new blocks to the blockchain \cite{saha2023protecting}, among other malicious behaviours, which would allow them to steal funds or manipulate the network to their advantage, causing confirmed transactions to be reversed, causing chaos, and undermining the integrity of the blockchain.
\subsubsection{Scheme Design}
In the honest branch, the miner who mined the new block and the miner who mined the previous block are likely to be the same miner, so the distribution of miners will reflect the historical ratio. In contrast, the malicious branch needs to concentrate the arithmetic power to attack, so the distribution of miners will be different from the normal distribution in history. Therefore, the distribution of miners in the historical block
can be used to differentiate between the honest and malicious branches.
 
In this scenario, if the miners of a particular branch have a lower representation in the historical block, then the weight of this branch in the total difficulty calculation will be reduced, so the attacker can only exceed the honest chain through longer branches or build up enough representation of the miners in the historical block to increase the credibility and then launch a 51\% attack.
\begin{enumerate}[label=\textbullet]
\item[] The workflow of this solution includes the following steps:
\item[(1).] Calculate the block generation frequency of miners.

For each miner $i$ in the historical window $W$, the formula for calculating the block generation frequency is:
\begin{align*}  
    r_i &=\frac{\text{\textit{Number of blocks mined by miner} } i}{\text{\textit{Total number of blocks in the history window}}},\text{\textit{and} }\sum_{i=0}^n r_i=1  \\
\end{align*}
\item[(2).] Each block is signed using the miner's private key to prevent falsification of the miner's identity.

\item[(3).] Calculate the historical weighted difficulty HWD for each branch.
The formula for calculating the historical weighted difficulty HWD of miner k in branch b when a fork is detected is:
\begin{align*}  
    HWD_b=HW_b \cdot D_b=\sum_{k=1}^l r_k \cdot \sum_{k=1}^l d_k  
\end{align*}

    Where $r_k$ is the frequency of block generation in the history window, $d_k$ is the difficulty of the block $k$, and $l$ is the branch length
in degrees. If the same miner mines more than one block, it is only counted once. This prevents a single high-computing-power miner from monopolizing mining, which encourages decentralization while making attacks more difficult.
\item[(4).]Each node compares the HWD values of the two branches and chooses the branch with the larger HWD as the master chain.

The key to this mechanism for defending against a 51\% attack lies in the attacker's power requirements and the difficulty of controlling branches. Suppose the attacker launches the attack only by temporarily increasing the arithmetic power. In that case, the miners in the branch are new to the system. Their IR is very low, and the corresponding HWD is also very low, so the nodes on the original chain will not switch to the attacker's chain. The scheme structure diagram is shown in Figure \ref{fig:8}.

\begin{figure}[ht]
  \centering
\includegraphics[width=1.0\textwidth]{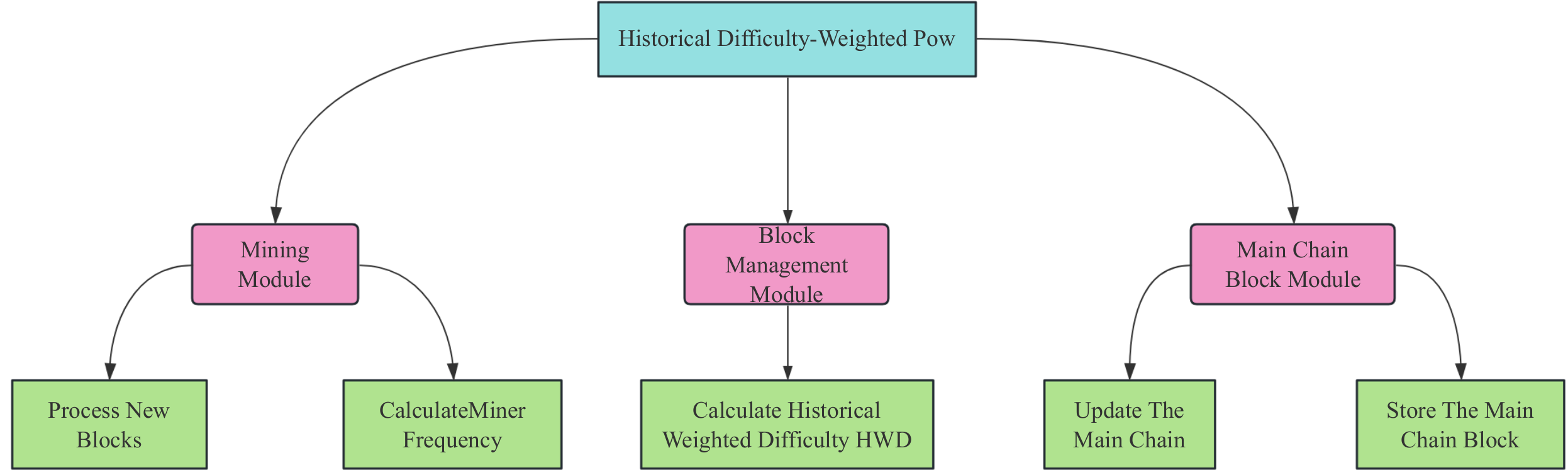}
  \caption{Structural Diagram of Historical Weighted Difficulty Design Scheme}  
   \label{fig:8}
\end{figure}

\end{enumerate}
\subsubsection{Result Comparison Analysis}
In this paper, we will compare the PoW consensus mechanism that introduces HWD with the ordinary PoW, in which the core function of the ordinary PoW consensus mechanism is shown in the Listing\ref{code1}, which includes the functions to determine whether mining can be successful or not and to carry out mining.
\begin{lstlisting}[language=Solidity, caption={Core Functionality of PoW Consensus Mechanisms}, label={code1}]
function isSuccess(uint256 nonce) public view returns(bool) { 
    return (uint256(keccak256(abi.encodePacked(nonce))) % difficulty == 0;
}

function mineBlock(uint256 nonce) public returns(bool) { 
    require(isSuccess(nonce), "Invalid Nonce");
    require(!usedNonces[nonce], "This nonce is already used!");
    
    lastTime = block.timestamp;
    minerBlockNumbers[msg.sender] += 1;
    usedNonces[nonce] = true;

    emit getNewBlock(msg.sender, difficulty, lastTime);
    return true;
}
\end{lstlisting}

Due to the situation, after the successful mining of blocks, the introduction of HWD enhanced the PoW consensus mechanism, no longer requiring the setting of a nonce value. instead, the default is to find an available nonce after successful block mining. The core functions in the HWD scheme include calculating the frequency of miners out of the block, calculating the value of the HWD, and updating the main chain. The code for the implementation of the core functions is shown in the Listing\ref{code2}-\ref{code3}

\begin{lstlisting}[language=Solidity, caption={Core Functionality of PoW Consensus Mechanisms}, label={code2}]
function calculateWinerBlockFrequency(address miner) internal view returns (uint256) {
    uint256 totalWinerBlockNumbers = 0; 
    uint256 totalBlockNumbers = historyBlocks.length; 
    
    if(totalBlockNumbers == 0) {
        return 0; 
    }
    
    for(uint256 i = 0; i < totalBlockNumbers; i++) {
        if(historyBlocks[i].miner == miner) {
            totalWinerBlockNumbers += 1;
        }
    }
    
    return totalWinerBlockNumbers * 1e18 / totalBlockNumbers;
}

function updateMainBranch() internal {
    if(mainBranch.length == 0) {

        mainBranch = historyBlocks;
    } else {
        uint256 mainBranchHMD = calculateHMD(mainBranch); 
        uint256 historyBlocksHMD = calculateHMD(historyBlocks); // 
        
        if(mainBranchHMD < historyBlocksHMD) {
            mainBranch = historyBlocks;
        }
    }
}
\end{lstlisting}

\begin{lstlisting}[language=Solidity, caption={Code Implementing the Calculation of HWD Values in the HWD Scheme}, label={code3}]
function calculateHMD(Block[] memory blocks) internal view returns (uint256) {
    uint256 hwd = 0;
    if (blocks.length < MinBlockNumbers) {
        return 0;
    }

    address[] memory uniqueMiners = new address[](blocks.length);
    uint256 uniqueCount = 0;

    for (uint256 i = 0; i < blocks.length; i++) {
        address miner = blocks[i].miner;
        uint256 difficulty = blocks[i].difficulty; 

        if (!minerUsed(uniqueMiners, miner, uniqueCount)) {
            uint256 minerBlockFrequency = calculateMinerBlockFrequency(miner);
            hwd += difficulty * minerBlockFrequency;
            uniqueMiners[uniqueCount] = miner;
            uniqueCount++;
        }
    }

    return hwd;
}

\end{lstlisting}

The PoW consensus mechanism can be vulnerable to a 51\%  attack if the malicious attacker's arithmetic power exceeds half of the entire system's. The code to simulate a 51\%  attack on PoW at the console is shown in Listing\ref{code4}, and the results of its contract execution are shown in Table\ref{result1}. When testing the PoW consensus mechanism with the introduction of HWD, the code to simulate a 51\% attack in the console is presented in Listing\ref{code6}, and the results of its execution are shown in Table \ref{result2} and \ref{result3}.

\begin{lstlisting}[caption={Code to Simulate 51\% Attack When Testing PoW}, label={code4}]

let pow = await pow.deployed(); 
let accounts = await web3.eth.getAccounts();

(async () => { 
    let validNonce = 0; 
    for (let i = 0; i < 2; i++) { 
        while (!(await pow.isSuccess(validNonce))) {
            validNonce++;
        }
        await pow.mineBlock(validNonce, {from: accounts[1]}); 
        console.log(`Mine block with nonce: ${validNonce}`);
        validNonce++;
    }
})();

(async () => { 
    let validNonce = 0; 
    for (let i = 0; i < 5; i++) { 
        while (!(await pow.isSuccess(validNonce))) {
            validNonce++;
        }
        try {
            await pow.mineBlock(validNonce, {from: accounts[2]}); 
            console.log(`Mine block with nonce: ${validNonce}`);
        } catch (error) {
            if (error.reason === "This nonce is already used!") {
                validNonce++;
                i--; 
                continue;
            } else {
                throw error;
            }
        }
        validNonce++;
    }
})();

let blockNumber1 = await pow.getMinerBlockNumbers(accounts[1]); 
let blockNumber2 = await pow.getMinerBlockNumbers(accounts[2]); 

console.log(`Miner 1 (${accounts[1]}) blocks: ${blockNumber1.toString()}`);
console.log(`Miner 2 (${accounts[2]}) blocks: ${blockNumber2.toString()}`);

let blockCount = await pow.getTotalBlockCount();
console.log(`Total blocks: ${blockCount.toString()}`);
\end{lstlisting}

\begin{table}[H]
  \centering  
  \caption{Testing Some of the Mainchain Blocks after the Attack During the HWD PoW}  
  \label{result1}  
 \begin{tabularx}{\textwidth}{X X X X}
 \toprule
  \textbf{Operational Phase} & \textbf{Number of Cycles} & \textbf{Successfully Mined Block Number}& \textbf{Account Block Statistics}\\
  \hline
  First Dig &2 Cycles &62,1148 &Hiner A (Account[i]) Blocks: 2\\
  Second Dig &5 Cycles &1188,1088,1072,1090 &Hiner A (Account[i]) Blocks: 5(Includes Error Handling)
 \\
  \hline
\end{tabularx}
\end{table}

\begin{lstlisting}[caption={Code to Simulate 51\% Attack When Testing HWD PoW}, label={code6}]
let hwdpow = await HWDPow.deployed();

historyBlocksCount = await hwdpow.getHistoryBlocksCount(); 

(async () => {for (let i = 0; i < 40; i++)  
    {let miner = (i % 2 == 0) ? accounts[3] : accounts[4];  
    await hwdpow.minedBlock(miner, 50, i, web3.utils.asciiToHex("signature"));
    console.log("Honest miners have mined 40 blocks.");}()); 

let mainBranch = await hwdpow.getMainBranch.call(); 
console.log("Main chain before attack:", mainBranch);

historyBlocksCount = await hwdpow.getHistoryBlocksCount();

(async () => {for (let i = 40; i < 100; i++)  
    {await hwdpow.minedBlock(accounts[5], 60, i, web3.utils.asciiToHex("malicious_signature"));
    console.log("Attacker has mined 60 blocks.");}()); 

mainBranch = await hwdpow.getMainBranch.call();  
console.log("Main chain after attack:", mainBranch);

historyBlocksCount = await hwdpow.getHistoryBlocksCount();
\end{lstlisting}

\begin{table}[H]
  \centering  
  \caption{A Portion of the Mainchain Block before the Attack was Performed While Testing HWD PoW}  
  \label{result2}  
 \begin{tabularx}{\textwidth}{l X X l}
 \toprule
  \textbf{Miner Address} & \textbf{Difficulty} & \textbf{Block Number}& \textbf{Signature/ID}\\
  \hline
  0x6F71Aa557bAD489795A26F89E5752f73E252002E & 50 &0 &0x7369676e6174757265\\
  0x275Ab69Ea0762C295199A2574b272Aab08f3c08d &50&1&0x7369676e6174757265
 \\
  \hline
\end{tabularx}
\end{table}

\begin{table}[H]
  \centering  
  \caption{Testing Some of the Mainchain Blocks after the Attack During the HWD PoW}  
  \label{result3}  
 \begin{tabularx}{\textwidth}{l l l X}
 \toprule
  \textbf{Miner Address} & \textbf{Difficulty} & \textbf{Block Number}& \textbf{Signature/ID}\\
  \hline
  06xf71Aa557bAD409795A26F09ES752f73E252002E & 50 &38 &0x7369676e6174757265\\
  0x275Av69EA0762C295199A2574b272AabB8f3c08d &50 &39 &0x7369676e6174757265
 \\
0x19Ac4Cc64e983857FF043A0915D32D7761c4425b &60 &40 &0x6d616c6963696f75735
\\
&&&f7369676e6174757265
 \\
  \hline
\end{tabularx}
\end{table}

The above test results show that in the traditional PoW mechanism, a strong attacker launching a 51\% attack can form a longer chain to reorganise the existing one, and the only way to resist such an attack is to increase the difficulty. The PoW mechanism with HWD still contains many blocks mined by normal miners on the main chain after a 51\% attack, which is capable of withstanding such an attack. After 40 blocks are mined by normal miners alternately, there are 40 blocks in the history block and 30 blocks in the main chain. Initially, the number of blocks in the main chain is 0. The first block mined will be added to the main chain directly, and if the number of blocks mined is less than the constant MinBlockNumbers, the HWD of the main chain's HWD and the historical blocks are both 0. When the number of blocks mined reaches MinBlockNumbers, by using Eq:

\begin{enumerate}[label=\textbullet]
\item[] 
\begin{align*}  
    r_i &=\frac{\text{\textit{Number of blocks mined by miner} } i}{\text{\textit{Total number of blocks in the history window}}}
\end{align*}
\item[]
The block frequency of miner 1 can be calculated as:
\begin{align*}
        r_1=\frac{15}{30}=0.5
\end{align*}
Miner 2 has a block out frequency of:
\begin{align*}
        r_2=\frac{15}{30}=0.5
\end{align*}
Therefore, the HWD value of the historical block is:
\begin{align*}
HWD=r_1 \cdot k_1 + r_2\cdot k_2=0.5\cdot 0.5 + 0.5\cdot 50=50>0
\end{align*}

Currently, the HWD of the historical blocks is larger than the HWD of the main chain, and the main chain will be updated to 30 blocks. Later, because miner one and miner two mine with the same difficulty and $\sum_{i=0}^n r_i=1 $, when they mine 40 blocks, the HWD of the historical blocks is 50, and the main chain is no longer updated. When the malicious miner mines the 41st block, the block frequency of miner 1 is:

\begin{align*}
        r_1=\frac{20}{41}=0.4878
\end{align*}
Miner 2 has a block out frequency of:
\begin{align*}
        r_2=\frac{20}{41}=0.4878
\end{align*}
The block frequency of the malicious miner is:
\begin{align*}
        r_3=\frac{1}{41}=0.02439
\end{align*}
Therefore, the HWD value of the historical block is:
\begin{align*}
        HWD=r_1 \cdot k_1+r_2 \cdot k_2+r_3 \cdot k_3=50.244>50
\end{align*}
The main chain will be updated to 41 blocks. However, the HWD of historical blocks will not exceed the HWD of the main chain due to the small growth of the HWD of historical blocks and errors, so the main chain will not be updated.
\end{enumerate}
\subsection{Double Spending Attack}
\subsubsection{Principle of Attack}
A double-spending attack is when the same digital currency is used two or more times, and since digital currencies are replicable, an attacker can send the same digital currency to several different recipients, destroying the irreversibility and consistency of transactions in the blockchain network\cite{das2024analysing}. The basic principle is to take advantage of the delays and loopholes in the consensus mechanism in the blockchain network by broadcasting two different transactions at the same time, one of which is confirmed while the other is hidden or replaced. An attacker can realize the double-spending attack by using attacks such as the direct double-spending attack, 51\% attack, the race attack, and the Finney attack.
\subsubsection{Scheme Design}
\begin{figure}[ht]
  \centering
\includegraphics[width=0.8\textwidth]{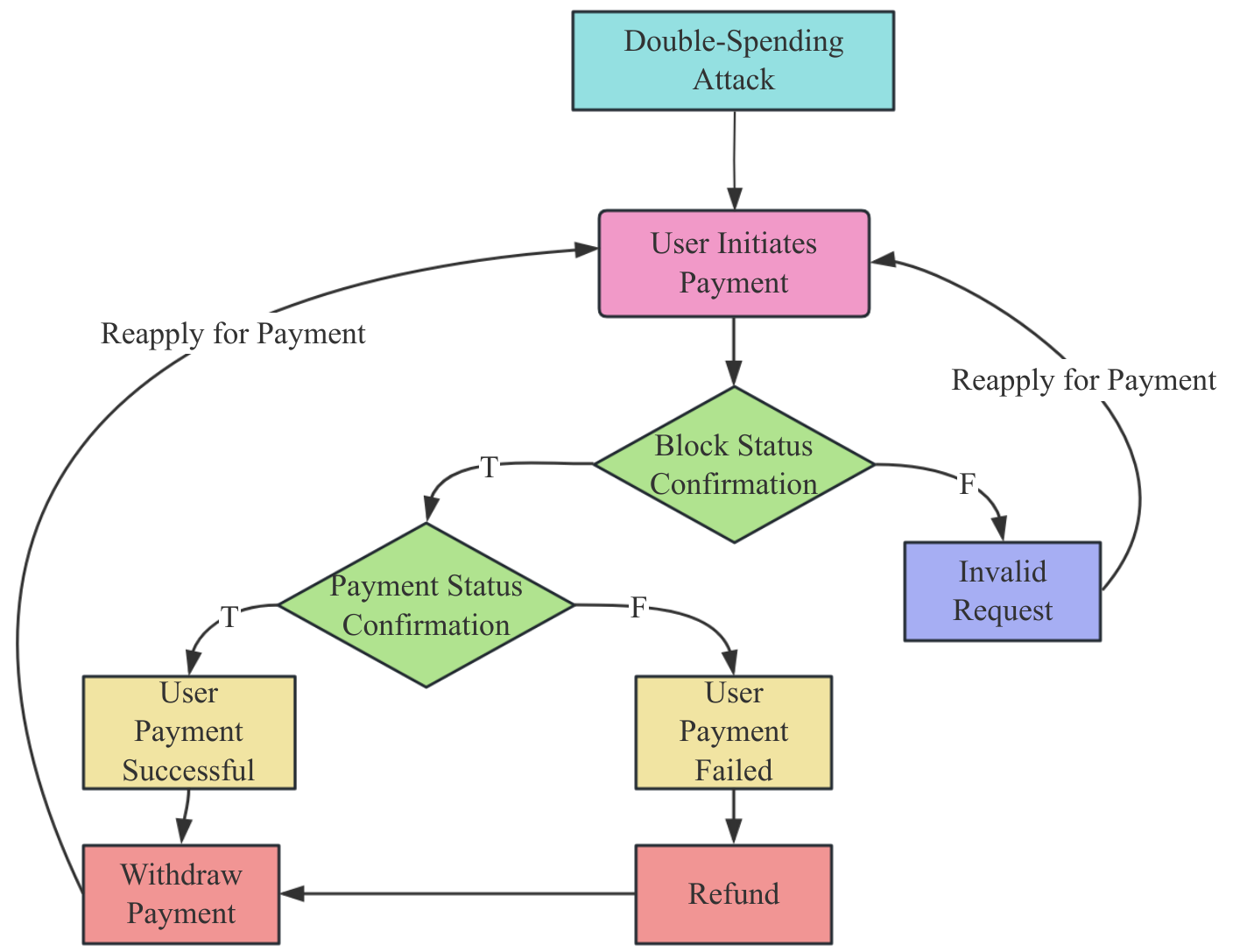}
  \caption{Structure Diagram of the Double-Spending Attack Design Scheme}  
   \label{fig:001}
\end{figure}

The Double-Spending attack can be prevented by adding a payment status and payment confirmation mechanism, i.e., the user's payment cannot be repeated until it is confirmed, and the structure of the design scheme is shown in Figure\ref{fig:001}. The user needs to wait for the blockchain network to confirm the block before the contract administrator confirms the payment, and only if the payment is not confirmed and the user fails to make the payment can the contract administrator confirm the payment, and only if the user fails to make the payment. Only if the payment is not confirmed and the user fails to make a refund will the user wait until the payment is confirmed or refunded before initiating a new payment request, limiting attackers' frequency of payment requests.

\subsubsection{Result Comparison Analysis}
In this paper, we will test and compare the effectiveness of the scheme designed to resist the double-spending attack with a common vulnerable scheme, where the main functions in the common contract include payment, confirmation of payment, and refund, and its specific code implementation is shown in Listing\ref{code2.1}.
\begin{lstlisting}[caption={Code Implementation of Payment, Payment Confirmation and Refund Functions for Common Contracts}, label={code2.1}]
function pay() external payable{ 
    require(msg.value > 0, "Amount must more than 0");
    balances[msg.sender] += msg.value;
}

function confirmPayment(address buyer, uint256 amount) external onlyOwner() { 
    require(balances[buyer] >= amount, "No enough money !");
    balances[buyer] -= amount;
}

function refund(address payable to, uint256 amount) external onlyOwner(){ 
    require(address(this).balance >= amount, "Don't have enough money in contract !");
    require(balances[to] >= amount, "Don't have enough money in account !");
    balances[to] -= amount;
    (bool success, ) = to.call(value:amount){"");
    require(success, "Refund failed !");
}
\end{lstlisting}

In the scheme that introduces the payment status and network confirmation, the payment status is an enumerated variable. Its code implementation is shown in Listing\ref{code2.2}, which limits the user's access to certain functions based on the payment status. The code implemented in the core functions of the scheme is also illustrated in Listing \ref{code2.3}-\ref{code2.5}.

\begin{lstlisting}[caption={Definition of Payment Status}, label={code2.2}]
enum PaymentStatus { 
    None, 
    TBC, 
    Confirmed, 
    Refunded 
}

struct Payment { 
    uint256 amount;
    uint256 blockNumber;
    PaymentStatus status;
    uint256 timestamp;
    bytes32 payHash;
\end{lstlisting}
\begin{lstlisting}[caption={Code Implementation of the Payment Function}, label={code2.3}]
function pay() external payable{ 
    require(msg.value > 0, "Amount must more than 0");
    require(payments[msg.sender].status == PaymentStatus.None || payments[msg.sender].status == PaymentStatus.Refunded,
    "Your last payment does not be confined or refund !"); 
    require(maxTBC > 0,"");

    bytes32 payHash = keccak256(abi.encodePacked(msg.sender, msg.value, block.number, block.timestamp));

    payments[msg.sender] = Payment({ 
        amount: msg.value,
        blockNumber:block.number,
        status:PaymentStatus.TBC, 
        timestamp:block.timestamp,
        payHash:payHash
    });

    maxTBC -= 1; 
    emit PaymentSend(msg.sender,msg.value);
\end{lstlisting}
\begin{lstlisting}[caption={Code Implementation of Payment Confirmation Function}, label={code2.4}]
function confirmPayment(address buyer) external onlyOwner(){ 
    Payment storage payment = payments[buyer];
    require(payment.amount > 0,"No money !");
    require(payment.status == PaymentStatus.TBC,
    "The payment don't need to confirm !"); 
    require(block.number >= payment.blockNumber + 24,
    "Payment don't be confirmed by network !"); 
    require(payment.payHash == keccak256(abi.encodePacked(
        buyer,
        payment.amount,
        payment.blockNumber,
        payment.timestamp)),
    "Payment hash mismatch!");
    
    payment.status = PaymentStatus.Confirmed;
    maxTBC += 1;
    delete payments[buyer]; 
    emit PaymentConfirmed(buyer, payment.amount);
}
\end{lstlisting}
\begin{lstlisting}[caption={Code Implementation of the Refund Function}, label={code2.5}]
function refund(address payable to) external onlyOwner noReentrancy(){ 
    Payment storage payment = payments[to];
    require(payment.amount > 0, "No money to refund");

    require(payment.status == PaymentStatus.TBC, "The payment can't refund !"); 
    payment.status = PaymentStatus.Refunded; 
    maxTBC += 1;

    (bool success, ) = to.call{value: payment.amount}(""); 
    require(success, "Refund failed !");

    emit PaymentRefunded(to, payment.amount);
}
\end{lstlisting}
For the ordinary vulnerable contract, the code to simulate an attack on it at the console is shown in Listing\ref{code2.6}. For the scheme that introduces the payment status and network confirmation mechanism, the code that simulates a double-spending attack on it at the console is identical to that of an ordinary contract. The effect of this simulated attack against it is shown in Table\ref{result2.1}.
\begin{lstlisting}[caption={Code to Simulate a Double-Spending Attack When Testing a Normal Contract}, label={code2.6}]
let Vbank = await Vulnerable.deployed();
let accounts = await web3.eth.getAccounts();

await Vbank.pay({from: accounts[1], value: web3.utils.toWei("1", "ether")}); 
await Vbank.pay({from: accounts[1], value: web3.utils.toWei("1", "ether")}); 

balance = await Vbank.getBalance(accounts[1]);
console.log("Balance:", balance.toString());

await Vbank.confirmPayment(accounts[1], web3.utils.toWei("1", "ether"), {from: accounts[0]}); 

await Vbank.refund(accounts[1], web3.utils.toWei("1", "ether"), {from: accounts[0]}); 
\end{lstlisting}

\begin{table}[H]
  \centering  
  \caption{Testing the Effect of Simulating a Double-Spending Attack When Improving a Contract}  
  \label{result2.1}  
 \begin{tabularx}{\textwidth}{X l X X}
 \toprule
\textbf{Step} & \textbf{Actions } & \textbf{Transaction Status } & \textbf{Block} \\
  \hline
Initial Payment&HBank.psy()&Success&Block:46\\
Duplicate Payment&HBank.psy()&Failed&Error: Revert Your Last Payment Does Not Be Confined or Refund \!\\
 Initial Confirmation& ConfirmPayment(Accounts[2]) &Failed&Error: Revert Payment Don't Be Confirmed By Network \!\\
 Block Generation&Cyclic Send Short Transactions&Failed&	Generate 24 Empty Blocks\\
 Final Confirmation&confirmPayment(Accounts[2])&Success&Block:71\\
  \hline
\end{tabularx}
\end{table}
As a result of the above attack, the standard scheme does not restrict the payment status. It can still initiate another payment before confirming the first, resulting in a double-spending attack, as the user initiates a payment again while an unconfirmed payment transaction is still in progress. In the improved scheme, the attacker attempts to initiate a payment again after the initial payment attempt. However, the payment initiation fails because the attacker's payment status changes to 'Pending Confirmation' after initiating the first payment. Since the network did not mine more than 24 blocks after initiating the payment, it failed to confirm the payment, and the administrator confirms that the payment failed. After the network has mined 24 blocks, the network confirms its success. The administrator then confirms the payment success again, and the user's payment status returns to the empty state None, with a payment amount of 0. As a result, the attacker fails to attempt a refund.
\subsection{Reentrancy Attack}
\subsubsection{Principle of Attack}
A reentrancy attack is a common vulnerability in smart contracts. the essence is that the contract function fails to handle the contract state change properly. The attacker takes advantage of the gap between the contract execution logic and the state update out of synchronization, inserts the attack code into the contract execution process, and causes the contract to execute the same operation multiple times through recursive calls before the contract is completed, to enable the attacker to obtain illegal benefits. The attacker usually utilizes the callback functions such as $fallback()$ or $receive()$ of the contract to call target functions such as $withdraw()$ after depositing funds into the contract, and maliciously triggers the recursive call of the target function during the transfer of funds, resulting in the funds being withdrawn multiple times.
\subsubsection{Scheme Design}
\begin{figure}[ht]
  \centering
\includegraphics[width=0.8\textwidth]{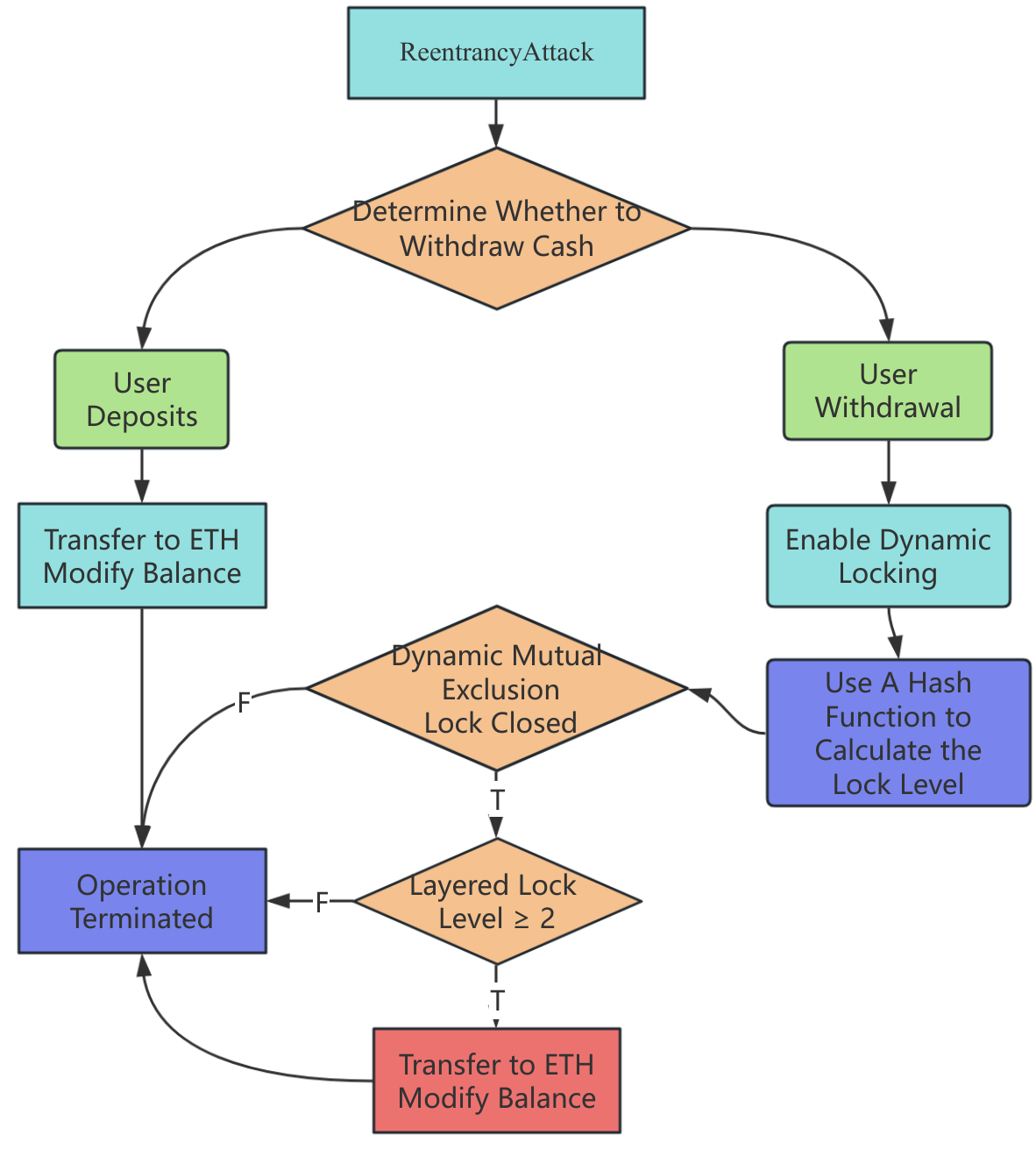}
  \caption{Reentrancy Attack Design Scheme Structure Diagram}  
   \label{fig:9}
\end{figure}

Mutual exclusion locks can be extended to prevent reentrancy attacks by combining dynamic mutual exclusion locks and hierarchical mutual exclusion locks \cite{kaur2024hybrid}. The structure of the design scheme is illustrated in Figure\ref{fig:9}. This mechanism assigns a unique mutex lock to each user account based on dynamic attributes, such as account balance, transaction history, or user activity. Then, different levels of hierarchical mutex locks are defined based on the criticality or sensitivity of the operation. The user's hierarchical mutex lock level should be checked before performing a critical operation, as higher-level locks may restrict access to more critical functions. The operation is executed only if both the user's dynamic mutex locks and hierarchical locks allow it, and both of the user's mutex locks are released upon completion of the operation.

Use $D_u$ to denote the dynamic mutex lock status of user $u$. When $D_u$ = 0, the user's dynamic mutex lock is not activated. $D_u$ = 1 means the user's dynamic mutex lock is activated. Use $H_u$ to denote the hierarchical lock level of user u. When $H_u$ = 0, the user has no hierarchical lock. when $H_u$ = 0 ($n\neq 0$), the user has different hierarchical lock levels. The state of the combined mutex lock $M_u$ of user $u$ is a combination of a dynamic mutex lock and a hierarchical mutex lock, which is defined as follows:
\begin{align*}
        M_u=D_u \cdot H_u
\end{align*}

When $D_u=0$, dynamic mutex locking is not activated and is not used regardless of the hierarchical locking level. when $D_u=1 $ and $H_u=0$, dynamic mutex locking is enabled but there is no hierarchical locking. and when $D_u=1$ and $H_u$=$n$ ($n\neq 0$), dynamic mutex locking is enabled and there is a specified hierarchical locking level, in contrast, the activation of a combinatorial mutex lock requires the setting of a hierarchical locking level s required for a critical operation, and the execution of the operation is allowed only if $D_u=1$ and $H_u$ $\geq$ $s$, when $M_u=1$, and only if $M_u=1$.

\subsubsection{Result Comparison Analysis}
In addition to implementing the improved scheme that introduces combined mutual exclusion locks and hierarchical locks, this paper will also implement a reentry attack using a simple contract to compare and analyse the results of the simulated execution of reentry attacks for both. Ordinary banking contract allows users to deposit and withdraw money as shown in Listing\ref{code3.1}, the improved scheme utilizes modifiers to prevent reentrant attacks by combining mutual exclusion locks and hierarchical locks, its code implementation is shown in Listing\ref{code3.2}, in addition, the attacking contract will utilize the receive() function to recursively call the withdraw() function to perform the reentrant attack, its core code implementation is shown in Listing\ref{code3.3}.
\begin{lstlisting}[caption={Code Implementation of the Main Functions of a Common Banking Contract}, label={code3.1}]
function deposit() external payable {
    require(msg.value > 0, "Amount must more than 0 !");
    balances[msg.sender] += msg.value;
}

function withdraw() public {
    if(balances[msg.sender] != 0) {
        (bool success, ) = msg.sender.call{value: balances[msg.sender]}("");
        require(success, "withdraw failed !");
        balances[msg.sender] = 0; 
    }
}
}
\end{lstlisting}
\begin{lstlisting}[caption={Improve Core Code Implementation in Banking Contracts}, label={code3.2}]
modifier nOReentranty() { 
    require(!dynamicMutex[msg.sender], "DynamicMutex is Locked !");
    dynamicMutex[msg.sender] = true;
    levelLock[msg.sender] = generateLevelLock(msg.sender); 
    _;
    dynamicMutex[msg.sender] = false; 
}

function generateLevelLock(address user) internal view returns (uint256) {
    bytes32 hash = keccak256(abi.encodePacked(user, balances(user)));
    uint256 level = uint256(hash) % 10; 
    return level;
}

function deposit() public payable { 
    require(msg.value > 0, "Amount must more than 0 !");
    balances[msg.sender] += msg.value;
}

function withdraw() public nOReentranty() {
    require(balances[msg.sender] > 0, "No money in account !");
    require(levelLock[msg.sender] >= 0, "Level lock is Locked"); 

    uint256 amount = balances[msg.sender];
    balances[msg.sender] = 0; 

    (bool success, ) = payable(msg.sender).call{value: amount}("");
    require(success, "withdraw failed !");
}
\end{lstlisting}

\begin{lstlisting}[caption={Code for Implementing the Main Functionality of the Contract for Performing Reentrancy Attacks}, label={code3.3}]
function depositToBank() public payable { 
    Bank(Contract).deposit{value: msg.value}();
}

function withdrawFromBank() public { 
    Bank(Contract).withdraw();
}

receive() external payable {
    if (Contract.balance >= 1 ether) { 
        Bank(Contract).withdraw();
    }
}
\end{lstlisting}

Since the reentrant attack is realised through recursion, the attack can be initiated by simply calling the attack function in the attack contract on the console. The specific execution effect of the reentrant attack simulated in the console for an ordinary banking contract is shown in Table \ref{result3.1}, and the resistance effect of the reentrant attack in the console for the improved scheme is shown in Table\ref{result3.2}.

\begin{table}[H]
  \centering  
  \caption{Testing the Effect of Simulated Reentrancy Attacks During Normal Contracts}  
  \label{result3.1}  
 \begin{tabularx}{\textwidth}{l X X c}
 \toprule
\textbf{Step} & \textbf{Key Operation} & \textbf{Vbank Balance} & \textbf{Attack Contract Balance} \\
  \hline
Initial State&Query Balance&1.0 ETH&\textbf{---}\\
Deposit Attack&Deposit 1 ETH&1.1ETH&\textbf{---}\\
&Tx Hash: 0x9c751023a&&\\
&87c4b53e8f78852b40e6&&\\
&96ef0c32d00316bdbc&&\\
&993320f160cb27a95&&\\
&Block:92&&\\
Withdrawal Attack&	Full Withdrawal&0ETH&\textbf{---}\\
&Tx Hash:0x8ffb38bdc02&&\\
&420a2aeb3409839c8fa3&&\\
&587e2ff34733a1e30238&&\\
&38a7a84876e14&&\\
&Block:93&&\\
Final State&Query Balance&0 ETH& 1.1ETH\\
  \hline
\end{tabularx}
\end{table}
\begin{table}[H]
  \centering  
  \caption{Results of Simulated Reentrancy Attacks When Testing Improved Contracts}  
  \label{result3.2}  
 \begin{tabularx}{\textwidth}{l X X c}
 \toprule
\textbf{Step} & \textbf{Key Operation} & \textbf{Ibank Balance} & \textbf{Result} \\
  \hline
Initial State&Query Balance&1.0 ETH&---\\
Deposit Attack&Deposit 1 ETH&1.1ETH&Success\\
&Tx Hash: 0x9c751023a&&\\
&87c4b53e8f78852b40e6&&\\
&96ef0c32d00316bdbc&&\\
&993320f160cb27a95&&\\
&Block:100&&\\
Withdrawal Attack&	Full Withdrawal&0ETH&Failed\\
Final State&Query Balance& 1.1ETH& Unchanged\\
  \hline
\end{tabularx}
\end{table}

In the vulnerable contract, the initial balance is 1.0 ETH. After a successful deposit of 1 ETH, the contract balance becomes 1.1 ETH. Due to the reentrancy vulnerability (transfer-before-state-update), the attacker can recursively call the withdraw function via the fallback mechanism and drain the entire contract balance of 1.1 ETH before the state update is executed, the attack contract triggers a re-entry attack by calling the function $withdrawFromBank()$: the vulnerable contract adopts the logic of "transferring money first and then updating the balance", the first time the transfer of $1 ETH$ triggers the $receive()$ callback function of the attack contract, which takes advantage of the loophole that the bank balance has not been updated (still $1.0 ETH$) to make a recursive call until the bank balance is zeroed out. The callback utilizes the loophole that the bank balance is not updated (there are still $1.0 ETH$) to achieve recursive calls, which are executed in a loop until the bank balance reaches zero. In the end, the attack contract successfully steals all 1.1 ETH, which verifies the effectiveness of the reentrancy attack. Improvements are made by using a mutex locking mechanism: adding the $noReentrancy$ modifier ensures that the key operation "updates the balance before transferring money". When the withdraw function is called, the dynamic mutex lock $dynamicMutex$ is immediately locked. The balance of the attacking contract is prioritised to be zeroed out before the transfer operation is performed. At this time, although the attack contract receives a callback, it attempts to call the withdraw function again. However, because the account balance has been cleared and the mutex lock is locked, the reentrancy request is rejected, triggering a transaction rollback and returning an "withdrawal failed" error. Ultimately, the bank balance remains unchanged at 1.1 ETH, confirming that the program can effectively block the reentrancy attack path and protect the security of the funds.

\subsection{Reply Attack}
\subsubsection{Principle of Attack}
Replay attacks \cite{alharby2017blockchain} are malicious or fraudulent network attacks that involve the repetition or delay of valid data. Due to the determinism and transparency of transactions in blockchain systems, attackers can intercept legitimate data transmissions without authorisation and rebroadcast transaction data that has already been legally signed and successfully executed by users to the blockchain network. This causes transactions to be executed multiple times in different states on the same chain or different chains, resulting in the loss of funds or resources.
\subsubsection{Scheme Design}
\begin{figure}[ht]
  \centering
\includegraphics[width=0.8\textwidth]{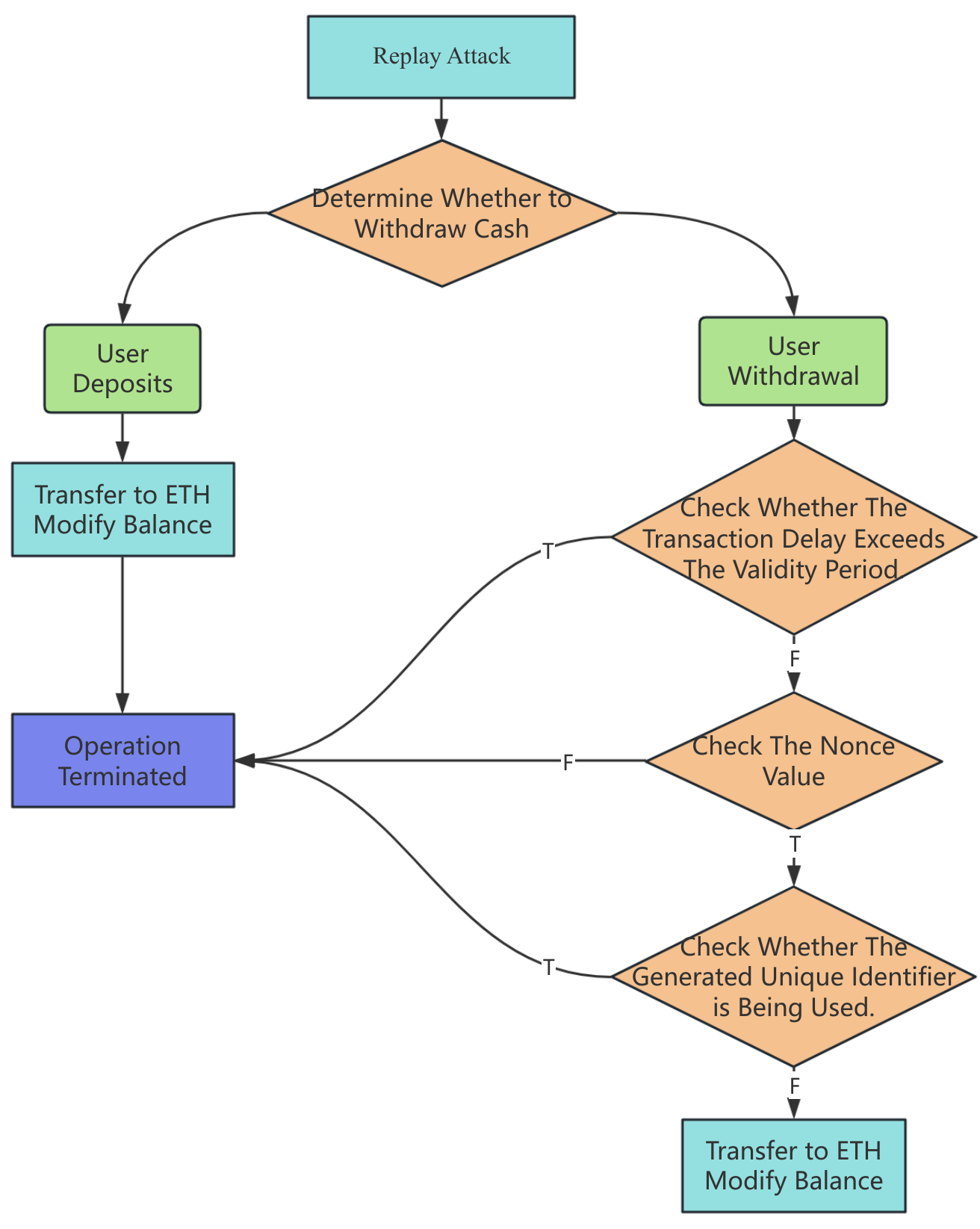}
  \caption{Structure Diagram of Replay Attack Design Scheme}  
   \label{fig:10}
\end{figure}

By introducing a nonce value and transaction validity period, it is possible to ensure that each transaction can only be executed once. Specifically, each user must carry a unique nonce value when conducting each transaction, and a unique identifier is generated through a hash function to ensure that transactions cannot be executed repeatedly, thereby preventing replay attackers from maliciously submitting old transactions multiple times within the network. Additionally, transactions must be completed within the validity period. transactions that exceed the time limit will be deemed invalid, preventing malicious attackers from repeatedly submitting transactions over time. The structural diagram of the design scheme is shown in Figure\ref{fig:10}.

\subsubsection{Result Comparison Analysis}
In the results analysis, this paper compares the resistance effects of the proposed scheme and a standard contract by simulating the same replay attack and analysing the results after the replay attack. The standard contract only includes deposit and withdrawal functions, and its main functional code is shown in Listing \ref{code4.1}. The improved contract introduces a nonce value and a transaction validity period, and its core functional code is shown in Listing \ref{code4.2}.

\begin{lstlisting}[caption={Code Implementation of the Main Functions of a Standard Contract}, label={code4.1}]
function deposit() external payable {
    require(msg.value > 0, "Amount must more than 0 !").
    balances[msg.sender] += msg.value;

    emit Deposit(msg.sender, msg.value);
}

function withdraw(uint256 amount) external {
    require(balances[msg.sender] >= amount, "No enough money in account");
    balances[msg.sender] -= amount;

    (bool success, ) = payable(msg.sender).call{value: amount}("");
    require(success, "Withdraw failed !");

    emit Withdraw(msg.sender, amount);
}
\end{lstlisting}

\begin{lstlisting}[caption={Code Implementation of the Main Functions of the Improved Banking Contract}, label={code4.2}]
function deposit() external payable {
    require(msg.value > 0, "Amount must more than 0!");
    balances[msg.sender] += msg.value; 
    emit Deposit(msg.sender, msg.value);
}

function withdraw(uint256 amount, uint256 nonce, uint256 timestamp) external {
    require(balances[msg.sender] >= amount, "No enough money in account !");
    require(block.timestamp <= timestamp + timeValid, "The time is overed !");

    bytes32 txHash = keccak256(abi.encodePacked(msg.sender, amount, nonce, timestamp)); 
    require(nonces[msg.sender] == nonce, "Invalid nonce !"); 
    require(!usedTxHashes[txHash], "Transaction already used !"); 

    nonces[msg.sender] += 1;
    balances[msg.sender] -= amount;
    lastTime[msg.sender] = block.timestamp;
    
    (bool success, ) = payable(msg.sender).call{value: amount}("");
    require(success, "withdraw failed !"); 

    usedTxHashes[txHash] = true; 
    emit Withdraw(msg.sender, amount, timestamp); 
}
\end{lstlisting}

Replay attacks obtain information about transactions that users have made through network monitoring and data analysis, then reintroduce this information into the network. The code for implementing a replay attack through network monitoring on the console is shown in Table\ref{code4.3}. The results of a replay attack on a standard contract are shown in Listing\ref{result4.1}, and the execution results of a simulated replay attack on an improved contract are shown in Table\ref{result4.2}.

\begin{lstlisting}[caption={Code for Simulating Replay Attacks When Testing Standard Contracts}, label={code4.3}]
await Vbank.deposit({ from: accounts[1], value: web3.utils.toWei("5", "ether") }); 

let balance = await Vbank.balances(accounts[1]);
console.log("Initial victim balance:", web3.utils.fromWei(balance.toString(), "ether"))
await Vbank.withdraw(web3.utils.toWei("2", "ether"), { from: accounts[1] }); 

balance = await Vbank.balances(accounts[1]);
console.log("Victim balance after withdraw:", web3.utils.fromWei(balance.toString(), "ether"))

let lastHash = (await web3.eth.getBlock("latest")).transactions[0]; 
console.log("Transaction hash:", lastHash);

let transaction = await web3.eth.getTransaction(lastHash); 
console.log("Transaction information:", transaction);

await web3.eth.sendTransaction({
    from: transaction.from, 
    to: transaction.to, 
    value: transaction.value, 
    gas: transaction.gas, 
    gasPrice: transaction.gasPrice, 
    data: transaction.input
});

balance = await Vbank.balances(accounts[1]); 
console.log("Final victim balance:", web3.utils.fromWei(balance.toString(), "ether"))
\end{lstlisting}

\begin{table}[H]
  \centering  
  \caption{Results of Simulating Replay Attacks When Testing Standard Contracts
  }  
  \label{result4.1}  
 \begin{tabularx}{\textwidth}{X X X X X}
 \toprule
\textbf{Step} & \textbf{Operation} & \textbf{Value} & \textbf{Transaction Status}&\textbf{Balance Changes} \\
  \hline
Initial Deposit&Vbank.Deposit()&Deposit 5 ETH&Success&\textbf{--}\\
Opening Balance&Check Balance&\textbf{--}&\textbf{--}&5 ETH\\ 
Withdrawal Operation&Vbank.Withdraw()&Withdraw 2 ETH&Success&\textbf{--}\\
Balance After Withdrawal&Check Balance&\textbf{--}&\textbf{--}&3ETH\\
Transaction Replay&Replay Last Trade&Copy Transaction Parameters&\textbf{--}&\textbf{--}\\
Final State&Check Balance&\textbf{--}&\textbf{--}&1ETH\\
  \hline
\end{tabularx}
\end{table}

\begin{table}[H]
  \centering  
  \caption{Code for Simulating Replay Attacks When Testing Contract Improvements
  }  
  \label{result4.2}  
 \begin{tabularx}{\textwidth}{X X X X X}
 \toprule
\textbf{Step} & \textbf{Operation} & \textbf{Value} & \textbf{Transaction Status}&\textbf{Balance Changes} \\
  \hline
Initial Deposit&Vbank.Deposit()&Deposit 5 ETH&Success&\textbf{--}\\
Opening Balance&Check Balance&\textbf{--}&\textbf{--}&5 ETH\\
Withdrawal Operation&Vbank.Withdraw()&Withdraw 2 ETH&Success&\textbf{--}\\
Balance After Withdrawal&Check Balance&\textbf{--}&\textbf{--}&3ETH\\
Transaction Replay&Replay Last Trade&Copy Transaction Parameters&Failed&\textbf{--}\\
Final State&Check Balance&\textbf{--}&\textbf{--}&3ETH\\
  \hline
\end{tabularx}
\end{table}
\subsection{Sybil Attack}
\subsubsection{Principle of Attack}
A Sybil attack is a typical identity spoofing pattern in self-organising network environments. Attackers infiltrate the system by spoofing multiple virtual identities, leveraging the scale advantage of node identities to gain control over resources, thereby achieving resource exploitation, consensus disruption, and other means to obtain illegal gains. Sybil Attacks can disrupt resource-sharing mechanisms in $P2P$ networks. They may consume the computational resources of normal nodes to form denial-of-service attacks. In extreme cases, they could even lead to the seizure of network control, resulting in the entire network service failing.
\subsubsection{Scheme Design}
\begin{figure}[ht]
  \centering
\includegraphics[width=0.8\textwidth]{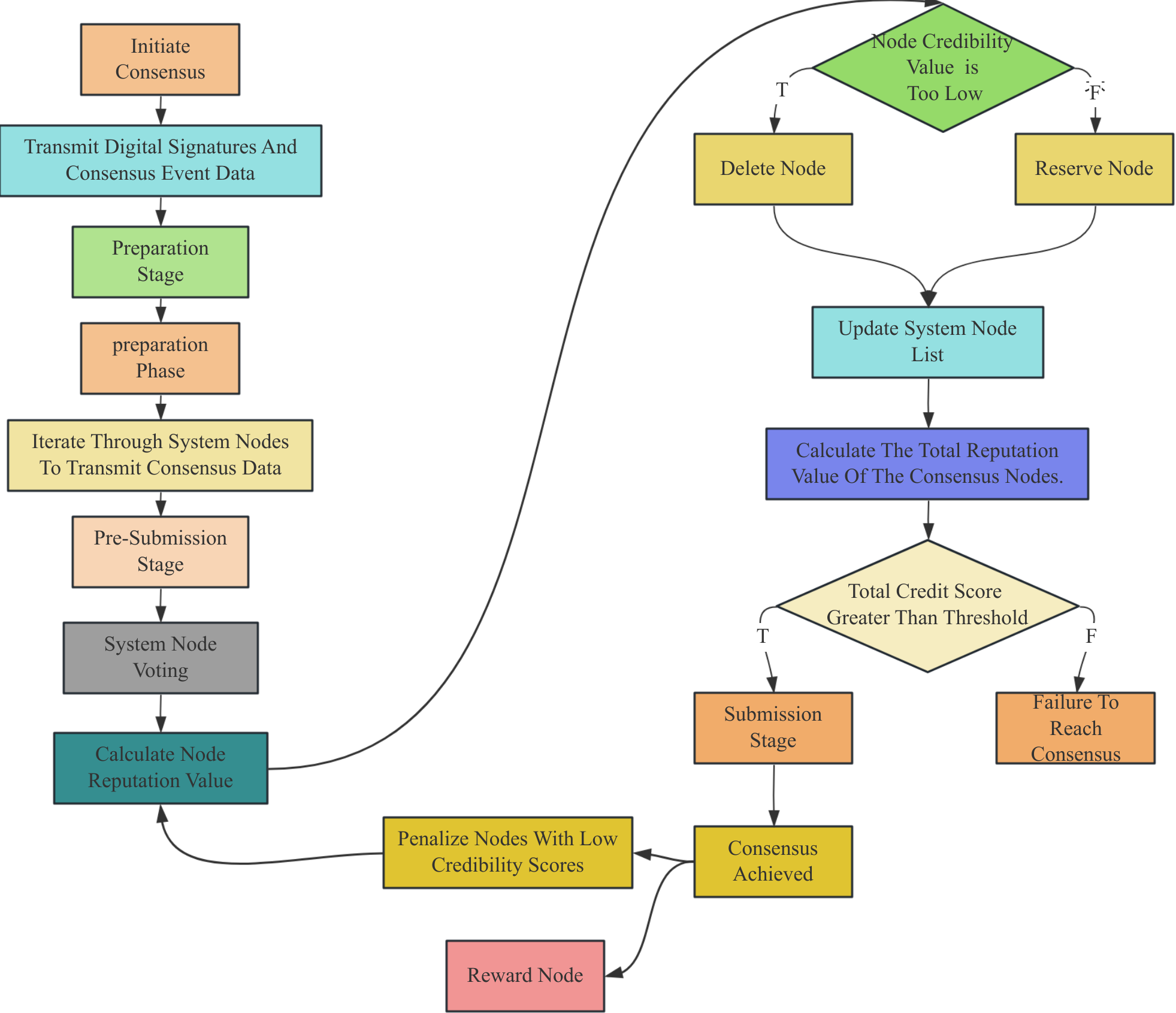}
  \caption{Structure Diagram of the Sybil Attack Design Scheme}  
   \label{fig:11}
\end{figure}

Based on the $PBFT$ consensus mechanism, a reputation value is assigned to each node \cite{wang2023defense}. The reputation value is determined by evaluating the information provided by each consensus node. This method can detect and remove attacking nodes while rewarding well-behaved nodes with higher reputation and greater influence during the consensus process. During the consensus process, node $i$ only updates its consensus state when the sum of reputation values $R_v$ from other consensus nodes is sufficient. The sum of the reputation values $R_v$ from other nodes is calculated by evaluating the reputation of each node during the consensus process and accumulating the reputation of conscientious nodes over time. This gradually reduces malicious nodes influence while increasing benign nodes' influence. The structural diagram of the design scheme is shown in Figure\ref{fig:11}.

Assume that the blockchain network is a directed network G($\varepsilon$,$E$). Here, $\varepsilon$ is a set of n nodes, and E is a set of directed edges with weights. In the blockchain network, the reputation value  $R_v$ obtained by node $i$ in $t$ 4 rounds of consensus is jointly determined by the reputation values of the nodes with which it interacts. The calculation formula is as follows:
\begin{align*}
      R_v(t)=R_i(t)+\sum_{\varepsilon_j=1}^n [R_j(t)-R_i(t)]\cdot (\varphi^T)_{ij}
\end{align*}

Where $R_i(t)$ represents the reputation value of node $i$ in consensus $t$, and $\varphi$ is the matrix composed of the reputation values of all nodes in the network. If $R_v$ is influenced by multiple consensus nodes, then the change in reputation value $\Delta$$R_i$($t$) received by node $i$ is the sum of the influences of all nodes, calculated as follows:
\begin{align*}
    \Delta R_i(t)=\sum_{\varepsilon_j=1}^n [R_j(t)-R_i(t)]\cdot (\varphi^T)_{ij}
\end{align*}
The credibility value $R_v$ from other nodes should be higher than the set threshold $R_{ts}$. The calculation formula is as follows:
\begin{align*}
R_{ts}=\frac{1}{N} \cdot (\lfloor \frac{N-1}{3}\rfloor \cdot2+1)
\end{align*}
\subsubsection{Result Comparison Analysis}
This paper will compare a standard voting contract with the proposed improvement scheme. The code implementation of the standard voting contract is shown in Listing \ref{code5.1}. In the improved contract, the voting result depends on the reputation value. A pre-submission phase is added to the $PBFT$ consensus mechanism to calculate the reputation value. In the implementation, the matrix $\varphi$ value is set to 1. Listing \ref{code5.2}-\ref{code5.3} shows the core functional code implementation.

\begin{lstlisting}[caption={Code Implementation of Ordinary Contracts}, label={code5.1}]
function vote(string memory elector) external { 
    require(!voted[msg.sender], "You already voted !"); 
    ticketNumber[elector] += 1;
    voted[msg.sender] = true;
}

function getTicketNumber(string memory elector) public view returns(uint256){ 
    return ticketNumber[elector];
}
\end{lstlisting}

\begin{lstlisting}[caption={Improve the Specific Code Pre-Submitted in the Contract}, label={code5.2}]
function precommit(bytes32 data) external { 
    require(status == Status.Prepare, "Status invalid --!");
    status = Status.PreCommit;
    uint256 sumRv = 0; 
    for(uint256 i = 0; i < nodeList.length; i++) { 
        address nodeAddress = nodeList[i];
        Node storage node = nodes[nodeAddress];
        int256 R_i_change = calculateReputationChange(nodeAddress); 
        if(R_i_change >= 0) {
            node.reputation += uint256(R_i_change);
        } else {
            node.reputation -= uint256(-R_i_change);
        }

        if(node.reputation <= 50) {
            deleteNode(nodeAddress);
        }

        if(node.isValid) {
            if(voted[nodeAddress]) {
                sumRv += node.reputation;
            }
        }
    }

    if (sumRv >= reputationThreshold) { 
        emit Precommit(msg.sender, blockHigh, true);
        commit(data);
    } else {
        emit Precommit(msg.sender, blockHigh, false);
    }
}
\end{lstlisting}
\begin{lstlisting}[caption={Improve the Threshold Calculation Implementation Code of the Contract.}, label={code5.3}]
function calculateThreshold(uint256 nodeNumber) internal pure returns(uint256) { 
    require(nodeNumber > 0, "No node!");
    uint256 f = (nodeNumber - 1) / 3; 
    return ((2 * f) + 1) * (50 / nodeNumber); 
}
\end{lstlisting}

The voting results of ordinary contracts are entirely dependent on node votes. The code for simulating a Sybil attack in the console is shown in Listing\ref{code5.6}, and the results of multiple Sybil attacks affecting consensus and resistance are shown in Table\ref{result5.7}.
\begin{lstlisting}[caption={Test Improvement Plan Code Simulating Sybil Attacks}, label={code5.6}]
totalNodes = await I_vote.getNodeNumber(); 
(async () => {
    for (let i = 0; i < totalNodes; i++) {
        let nodeAddr = await I_vote.nodeList(i); 
        let node = await I_vote.nodes(nodeAddr); 
        console.log(`Node: ${nodeAddr} Reputation: ${node.reputation}, isValid: ${node.isValid}`);
    }
})();

await I_vote.request(web3.utils.asciiToHex("Transaction")); 
await I_vote.vote(true, { from: accounts[1] }); 
await I_vote.vote(true, { from: accounts[2] }); 
await I_vote.vote(false, { from: accounts[3] }); 
await I_vote.vote(false, { from: accounts[4] }); 
await I_vote.vote(false, { from: accounts[5] }); 

await I_vote.preCommit(web3.utils.keccak256("Transaction")); 
(async () => {
    for (let i = 0; i < totalNodes; i++) {
        let nodeAddr = await I_vote.nodeList(i); 
        let node = await I_vote.nodes(nodeAddr); 
        console.log(`Node: ${nodeAddr} Reputation: ${node.reputation}, isValid: ${node.isValid}`);
    }
})();
\end{lstlisting}

\begin{table}[H]
  \centering  
  \caption{Changes in Node Reputation Values after Multiple Consensus Rounds
  }  
  \label{result5.7}  
 \begin{tabularx}{\textwidth}{X X X}
 \toprule
\textbf{Phase/Category} & \textbf{Description} & \textbf{Value/Result}  \\
  \hline
Initial State&Number of Nodes&5\\
&Initial Reputation Per Node	&100\\
&Total Initial Reputation	&500\\
Pre-Commit Phase&Reputation Change Amount&0 (No Change)\\
&Nodes Supportingconsensus (Honest)	&2\\
&Nodes Opposing Consensus (Malicious)&	3\\
&Pre-Commit Total Reputation (Supporters Only)&	200\\
&Threshold&30\\
&Consensus Result&Successful\\
Commit Phase	&Honest Node Reputation Change&+10\\
&Malicious Node Reputation Change (Penalty)	&-10 \\
Post-Commit State&Honest Node Reputation (2 Nodes)&110 \\
&Malicious Node Reputation (3 Nodes)	&90 \\
&Total Reputation &490\\
  \hline
\end{tabularx}
\end{table}
\subsection{Time-Bandit Attack}
\subsubsection{Principle of Attack}
A Time-Bandit Attack refers to an attacker exploiting the variability and inaccuracy of timestamps in a blockchain system to manipulate specific time-stamp-based mechanisms or conditions, thereby impacting the network or users. Timestamps in blockchain are typically set by miners when blocks are generated, so these timestamps are unreliable. Attackers can exploit this by manipulating timestamps to launch attacks. Many smart contracts rely on timestamps to determine the contract's state or execution conditions. If an attacker can manipulate timestamps, they can prematurely trigger specific functions in the contract or delay them, making it easier to gain an advantage \cite{cai2024assessing}.
\subsubsection{Scheme Design}
\begin{figure}[ht]
  \centering
\includegraphics[width=0.8\textwidth]{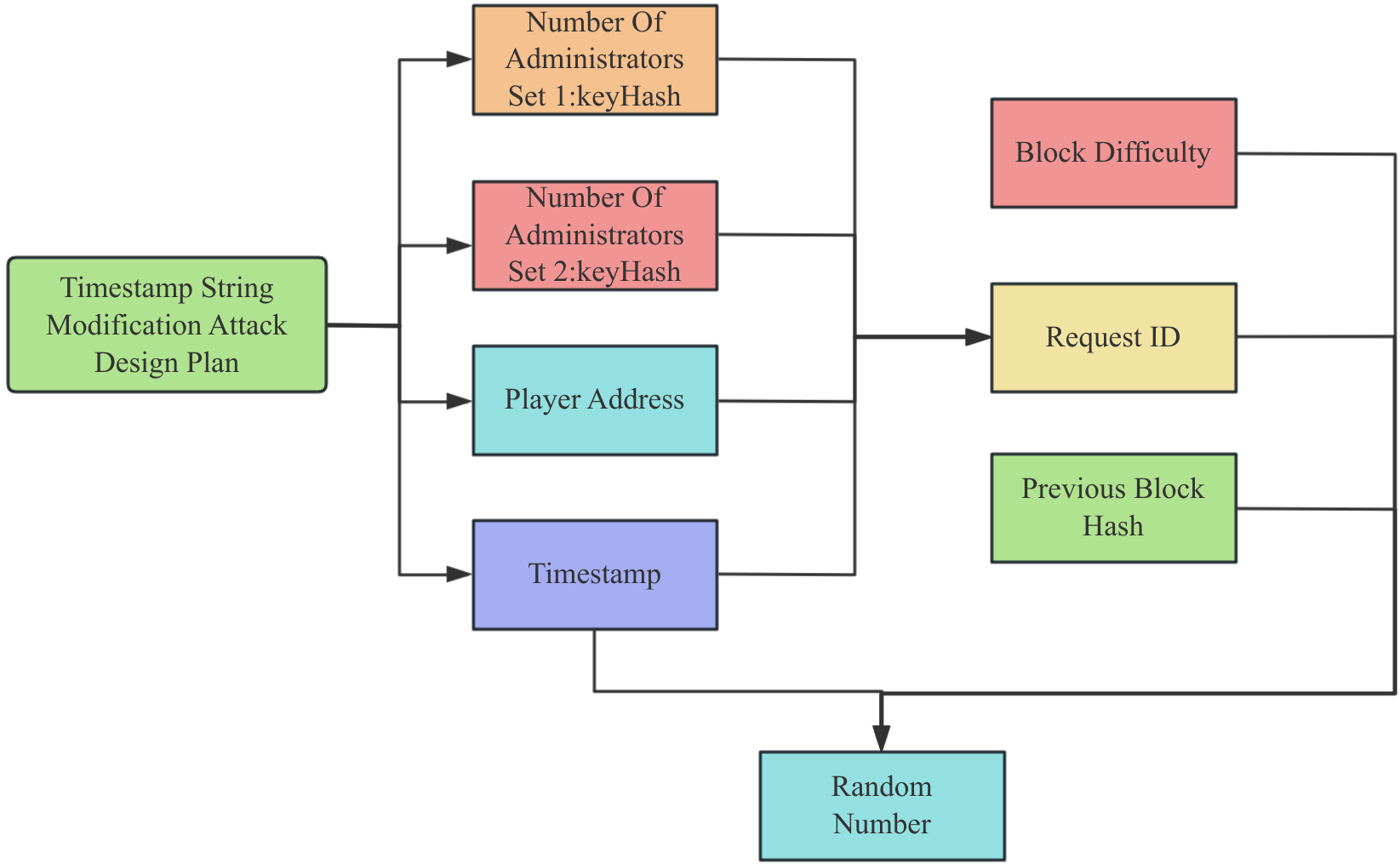}
  \caption{Structural Diagram of the Time-Bandit Attack Design Scheme}  
   \label{fig:12}
\end{figure}

Oracle \cite{koblitz2015random} can obtain accurate time data from external trusted sources and submit it to the blockchain, effectively avoiding the risk of inaccurate and tamperable timestamps on the chain. Chainlink VRF uses public-key cryptography to generate accurate random numbers and forms tamper-proof credentials through private-key signatures. Smart contracts verify the integrity of the signatures to ensure the reliability and accuracy of the random numbers, thereby resisting malicious attacks and ensuring the fairness of on-chain transactions. However, since $Chainlink VRF$ relies on external oracles to perform generation and verification in the network environment, this paper adopts an alternative hybrid random number generation scheme. This scheme not only relies on timestamps but also introduces other unpredictable parameters: first, the contract administrator sets parameters, block timestamps, and user addresses, which are then hashed to generate a “request $ID$” that is difficult for attackers to interfere with; subsequently, block information, timestamps, and this request $ID$ are combined to generate the final random number. This multi-dependent design (as shown in Figure\ref{fig:12}) ensures the unpredictability and security of random number generation.

\subsubsection{Result Comparison Analysis}
The proposed scheme will be compared with one that relies solely on block timestamps to generate random numbers. Both schemes are based on a lottery game that uses random numbers, where users deposit funds into the contract, the contract generates random numbers to determine if they have won, then executes the subsequent payment logic. The core code of the ordinary contract that produces random numbers solely based on timestamps is shown in Listing\ref{code6.1}. In the improved scheme, random number generation is not exclusively dependent on timestamps. Instead, it generates a request ID based on two numbers set by the contract administrator, the block timestamp, and the player's address. Then, it generates random numbers using the block hash, block timestamp, request ID, and block difficulty. The core functionality of the improved scheme is implemented as shown in Listing \ref{code6.2}.

\begin{lstlisting}[caption={Core Code of a Standard Contract}, label={code6.1}]
function play() external payable {
    require(msg.value > 0.1 ether, "Bet more money!");

    uint256 random = getRandom(block.timestamp) % 2;
    if(random == 1) {
        (bool success, ) = payable(msg.sender).call{value: msg.value * 2}("");
        require(success, "Transfer failed!");
        emit betResult(msg.sender, msg.value, true);
    } else {
        emit betResult(msg.sender, msg.value, false);
    }
}

function getRandom(uint timestamp) public pure returns(uint256) { /
    return uint256(keccak256(abi.encodePacked(timestamp)));
}

function withdraw(uint256 amount) external { 
    require(msg.sender == owner, "You can't use this function !");
    require(address(this).balance >= amount, "Not enough money in contract !");

    payable(owner).transfer(amount);
}
\end{lstlisting}
\begin{lstlisting}[caption={Improve the Code for Core Functions in the Contract.}, label={code6.2}]
function play() external payable {
    require(msg.value > 0.1 ether, "Bet more money !");
    Bets[msg.sender] = msg.value;

    bytes32 requestId = getRequestId(msg.sender); 
    uint256 random = getRandom(requestId); 
    if(random % 2 == 1) {
        (bool success, ) = payable(msg.sender).call{value: msg.value * 2}("");
        require(success, "Transfer failed!");
        emit betResult(msg.sender, msg.value, true);
    } else {
        emit betResult(msg.sender, msg.value, false);
    }
}

function getRequestId(address player) internal view returns (bytes32 requestId) {
    requestId = keccak256(abi.encodePacked(
        keyHash,
        number,
        block.timestamp,
        player));

function getRandom(bytes32 requestId) internal view returns(uint256 random) { 
    random = uint256(keccak256(abi.encodePacked(
        blockhash(block.number - 1),
        block.timestamp,
        block.difficulty,
        requestId));
}

function withdraw(uint256 amount) external onlyOwner { 
    require(address(this).balance >= amount, "Not enough money in contract !");
    payable(owner).transfer(amount);
}
\end{lstlisting}

When conducting a timestamp tampering attack, in addition to increasing the block timestamp, it is also necessary to mine blocks to make the modified timestamp effective. The code for conducting a timestamp tampering attack in the console is shown in Listing\ref{code6.3}, and the effect of improving the contract for conducting a timestamp tampering attack is shown in Table\ref{result6.1}.

\begin{lstlisting}[caption={Code for Simulating Timestamp Tampering Attacks When Testing Ordinary Contracts}, label={code6.3}]
await Vgame.sendTransaction({from: accounts[0], value: web3.utils.toWei("5", "ether")}); 
balance = await web3.eth.getBalance(Vgame.address); 
(async () => {
    while (true) {
        let block = await web3.eth.getBlock("latest");
        let timestamp = block.timestamp;
        let random = await Vgame.getRandom(timestamp);
        console.log(Current block timestamp:{timestamp},calculated random number:{random});
        
        if (BigInt(random) % 2n == 1n) {
            console.log('Find the right time stamp to place your bet');
            await Vgame.play({from: attacker, value: web3.utils.toWei("1", "ether")});
            console.log("Success!");
            break;
        }
        await new Promise((resolve, reject) => {
            web3.currentProvider.send(
                { jsonrpc: "2.0", method: "evm_increaseTime", params: [1], id: new Date().getTime() },
                (err, res) => (err ? reject(err) : resolve(res))
            );
        });
        await new Promise((resolve, reject) => {
            web3.currentProvider.send(
                { jsonrpc: "2.0", method: "evm_mine", params: [1], id: new Date().getTime() },
                (err, res) => (err ? reject(err) : resolve(res))
            );
        });
    }
})();
\end{lstlisting}

\begin{table}[H]
  \centering  
  \caption{Simulating the Effects of Timestamp Tampering Attacks When Testing Contract Improvements
  }  
  \label{result6.1}  
 \begin{tabularx}{\textwidth}{X X X}
 \toprule
\textbf{Steps}  & \textbf{Transaction/Result Information} & \textbf{Contract Balance}  \\
  \hline
Initial State&Number of Nodes&5\\
&Initial Reputation Per Node	&100\\
&Total Initial Reputation	&500\\
Pre-Commit Phase&Reputation Change Amount&0 (no change)\\
&Nodes Supportingconsensus (Honest)	&2\\
&Nodes Opposing Consensus (Malicious)&	3\\
&Pre-Commit Total Reputation (Supporters Only)&	200\\
&Threshold&30\\
&Consensus Result&Successful\\
Commit Phase	&Honest Node Reputation Change&+10\\
&Malicious Node Reputation Change (Penalty)	&-10 \\
Post-Commit State&Honest Node Reputation (2 Nodes)&110 \\
&Malicious Node Reputation (3 Nodes)	&90 \\
&Total Reputation &490\\
  \hline
\end{tabularx}
\end{table}
\section{CONCLUTION}
With its decentralised, transparent, immutable, and secure characteristics, blockchain technology has been widely adopted across various industries. However, as its applications expand, blockchain systems still face specific vulnerabilities in privacy protection, identity verification, and transaction security. This paper first explores the cryptographic techniques used in blockchain and their roles within the system. It then analyses the attack threats faced by each layer of the six-layer blockchain system architecture, explains the attack principles of these methods, and identifies defensive measures to mitigate or counter these attacks. Finally, it combines existing solutions and provides specific solutions to counter 51\% attacks, double-spending attacks, Re-entrancy attacks, replay attacks, Sybil attacks, and timestamp tampering attacks, thereby enhancing the blockchain system's resistance to attacks. Finally, it summarises the main evolutionary directions of blockchain technology to facilitate the construction of high-performance blockchain systems.
Looking ahead, the evolution of blockchain security protection faces dual challenges of technological innovation and expanding application scenarios. Future security technology research should focus on optimising core areas such as cryptographic techniques, peer-to-peer network architecture, consensus mechanisms, and privacy protection to build a more robust blockchain infrastructure. Only through the collaborative advancement and integration of cross-disciplinary technologies can blockchain systems effectively defend against complex network and algorithmic attacks, providing a solid foundation for the deep application of blockchain technology.
\bibliographystyle{plain}  
\bibliography{references}  

@misc{nakamoto2008bitcoin,
  title={Bitcoin: A Peer-to-Peer Electronic Cash System},
  author={Nakamoto, Satoshi},
  year={2008}
}

@article{shen2024survey,
  title={Survey on Blockchain Security Issues},
  author={Shen, Chuannian},
  journal={Computer Engineering and Science},
  volume={46},
  number={1},
  pages={46--62},
  year={2024}
}

@article{liu2024progress,
  title={Research Progress on Security Protection Technologies for Blockchain Systems},
  author={Liu, Aodi and Du, Xuehui and Wang, Na and others},
  journal={Chinese Journal of Computers},
  volume={47},
  number={3},
  pages={608--646},
  year={2024}
}

@article{tan2024improvement,
  title={Improvement of POW-based Blockchain Consensus Mechanism},
  author={Tan, Minsheng and Xu, Guoqing and Ding, Lin and others},
  journal={Computer Applications and Software},
  volume={41},
  number={11},
  pages={117--122},
  year={2024}
}

@article{jiang2024efficient,
  title={Efficient and Secure Improved DPoS Scheme},
  author={Jiang, Yibin and Wang, Xueming},
  journal={Computer and Digital Engineering},
  volume={52},
  number={10},
  pages={2996--3002},
  year={2024}
}

@phdthesis{xie2023research,
  title={Research on Security and Privacy Protection in Typical Blockchain Applications},
  author={Xie, Haomeng},
  school={Xidian University},
  year={2023},
  doi={10.27389/d.cnki.gxadu.2023.000301}
}

@article{xie2024spcex,
  title={SPCEX: Secure and Privacy-Preserving Cryptocurrency Exchange},
  author={Xie, Haomeng and Yan, Zheng},
  journal={IEEE Transactions on Dependable and Secure Computing},
  volume={21},
  number={5},
  pages={4404--4417},
  year={2024},
  doi={10.1109/TDSC.2024.3353541}
}

@inproceedings{yang2019effective,
  title={Effective Scheme against 51\% Attack on Proof-of-Work Blockchain with History Weighted Information},
  author={Yang, Xiaochen and Chen, Yu and Chen, Xiang},
  booktitle={IEEE International Conference on Blockchain},
  pages={261--265},
  year={2019},
  doi={10.1109/Blockchain.2019.00041}
}

@inproceedings{m2025efficient,
  title={Efficient Determent of Sybil Attacks in Blockchain},
  author={M, G. and Y, D. J and C, R. K and SK, S.},
  booktitle={International Conference on Multi-Agent Systems for Collaborative Intelligence},
  pages={268--272},
  year={2025},
  doi={10.1109/ICMSCI62561.2025.10894178}
}

@inproceedings{dey2018securing,
  title={Securing Majority-Attack in Blockchain Using Machine Learning and Algorithmic Game Theory: A Proof of Work},
  author={Dey, Soumya},
  booktitle={Computer Science and Electronic Engineering},
  pages={7--10},
  year={2018},
  doi={10.1109/CEEC.2018.8674185}
}

@article{yi2021efficient,
  title={An Efficient Blockchain Consensus Algorithm Based on Post-Quantum Threshold Signature},
  author={Yi, Haibo and Li, Yueping and Wang, Mei and Yan, Zengxian and Nie, Zhe},
  journal={Big Data Research},
  volume={26},
  pages={100268},
  year={2021},
  doi={10.1016/j.bdr.2021.100268}
}

@article{dai2022ddos,
  title={A DDoS-Attack Detection Method Oriented to the Blockchain Network Layer},
  author={Dai, Qinyu and Zhang, Bin and Dong, Shuqi},
  journal={Security and Communication Networks},
  volume={},
  year={2022}
}

@article{lu2024tree,
  title={Tree-ORAP: A Tree-Based Oblivious Random-Access Protocol for Privacy-Protected Blockchain},
  author={Lu, Yan and others},
  journal={IEEE Transactions on Services Computing},
  volume={17},
  number={3},
  pages={1252--1264},
  year={2024},
  doi={10.1109/TSC.2023.3347589}
}

@article{tian2021research,
  title={Research Progress on Attacks and Defense Technologies for Blockchain Systems},
  author={Tian, Guohua and Hu, Yunhan and Chen, Xiaofeng},
  journal={Journal of Software},
  volume={32},
  number={5},
  pages={1495--1525},
  year={2021},
  doi={10.13328/j.cnki.jos.006213}
}

@article{zhu2016security,
  title={Security Research on Key Blockchain Technologies},
  author={Zhu, Yan and Gan, Guohua and Deng, Di and others},
  journal={Information Security Research},
  volume={2},
  number={12},
  pages={1090--1097},
  year={2016}
}

@article{zhu2020research,
  title={Research Progress on Key Blockchain Technologies and Their Applications},
  author={Zhu, Jianming and Zhang, Quinnan and Gao, Sheng},
  journal={Journal of Taiyuan University of Technology},
  volume={51},
  number={3},
  pages={321--330},
  year={2020},
  doi={10.16355/j.cnki.issn1007-9432tyut.2020.03.001}
}

@inproceedings{saini2022blockchain,
  title={Blockchain and Cryptography},
  author={Saini, Kuldeep and Sharma, Shivani and Sarkar, Utpal},
  booktitle={2022 4th International Conference on Advances in Computing, Communication Control and Networking},
  pages={1863--1868},
  year={2022},
  doi={10.1109/ICAC3N56670.2022.10074345}
}

@inproceedings{ahmad2023study,
  title={Study of Cryptographic Techniques Adopted in Blockchain},
  author={Ahmad, Shabir and Arya, Sanjeev Kumar and Gupta, Sunil and Singh, Pankaj and Dwivedi, Sandeep Kumar},
  booktitle={4th International Conference on Intelligent Engineering and Management},
  pages={1--6},
  year={2023},
  doi={10.1109/ICIEM59379.2023.10166591}
}

@article{cai2021principles,
  title={Principles of Blockchain and Its Core Technologies},
  author={Cai, Xiaoqing and Deng, Yao and Zhang, Liang and others},
  journal={Chinese Journal of Computers},
  volume={44},
  number={1},
  pages={48},
  year={2021},
  doi={10.11897/SP.J.1016.2021.00084}
}

@article{raikwar2019sok,
  title={SoK of Used Cryptography in Blockchain},
  author={Raikwar, Mayank and Gligoroski, Danilo and Kralevska, Katina},
  journal={IEEE Access},
  volume={7},
  pages={148550--148575},
  year={2019},
  doi={10.1109/ACCESS.2019.2946983}
}

@inproceedings{lamba2013efficient,
  title={An Efficient Elliptic Curve Digital Signature Algorithm (ECDSA)},
  author={Lamba, Sanjay and Sharma, Manoj},
  booktitle={International Conference on Machine Intelligence and Research Advancement},
  pages={179--183},
  year={2013},
  doi={10.1109/ICMIRA.2013.41}
}

@incollection{rivest2006how,
  title={How to Leak a Secret: Theory and Applications of Ring Signatures},
  author={Rivest, Ronald and Shamir, Adi and Tauman, Yael},
  booktitle={Theoretical Computer Science},
  volume={3895},
  pages={164--186},
  year={2006},
  publisher={Springer}
}

@article{itakura1983public,
  title={A Public-Key Cryptosystem Suitable for Digital Multisignatures},
  author={Itakura, Keiichi and Nakamura, Kiyoshi},
  journal={NEC Research \& Development},
  volume={71},
  number={3},
  pages={1--8},
  year={1983}
}

@inproceedings{chaum1983blind,
  title={Blind Signatures for Untraceable Payments},
  author={Chaum, David},
  booktitle={Advances in Cryptology -- CRYPTO'82},
  pages={199--203},
  year={1983},
  doi={10.1007/978-1-4757-0602-4_18}
}

@incollection{desmedt1992shared,
  title={Shared Generation of Authenticators and Signatures},
  author={Desmedt, Yvo and Frankel, Yair},
  booktitle={Advances in Cryptology -- CRYPTO'91},
  pages={457--469},
  year={1992},
  publisher={Springer}
}

@inproceedings{sasson2014zerocash,
  title={Zerocash: Decentralized Anonymous Payments from Bitcoin},
  author={Ben Sasson, Eli and others},
  booktitle={IEEE Symposium on Security and Privacy},
  pages={459--474},
  year={2014},
  doi={10.1109/SP.2014.36}
}

@article{luong2023privacy,
  title={Privacy-Preserving Identity Management System on Blockchain Using Zk-SNARK},
  author={Luong, Duc A. and Park, Jong Hyuk},
  journal={IEEE Access},
  volume={11},
  pages={1840--1853},
  year={2023},
  doi={10.1109/ACCESS.2022.3233828}
}

@article{ben2018scalable,
  title={Scalable, transparent, and post-quantum secure computational integrity},
  author={Ben-Sasson, Eli and Bentov, Iddo and Horesh, Yinon and Riabzev, Michael},
  journal={Cryptology ePrint Archive},
  year={2018}
}

@inproceedings{bunz2018bulletproofs,
  title={Bulletproofs: Short Proofs for Confidential Transactions and More},
  author={Bünz, Benedikt and Bootle, Jonathan and Boneh, Dan and others},
  booktitle={IEEE Symposium on Security and Privacy},
  pages={315--334},
  year={2018},
  doi={10.1109/SP.2018.00020}
}

@inproceedings{maller2019sonic,
  title={Sonic: Zero-Knowledge SNARKs from Linear-Size Universal and Updatable Structured Reference Strings},
  author={Maller, Mary and Bowe, Sean and Kohlweiss, Markulf and Meiklejohn, Sarah},
  booktitle={Proceedings of the ACM SIGSAC Conference},
  pages={2111--2128},
  year={2019},
  doi={10.1145/3319535.3339817}
}

@article{yu2024research,
  title={Research on Quantum Algorithm Attacks on Bitcoin Blockchain},
  author={Yu, Jian and Yan, Fangxu and Wang, Jianhui and others},
  journal={Journal of Shenyang Normal University (Natural Science Edition)},
  volume={42},
  number={3},
  pages={222--227},
  year={2024}
}

@inproceedings{shor1994algorithms,
  title={Algorithms for Quantum Computation: Discrete Logarithms and Factoring},
  author={Shor, Peter W.},
  booktitle={Proceedings 35th Annual Symposium on Foundations of Computer Science},
  pages={124--134},
  year={1994},
  doi={10.1109/SFCS.1994.365700}
}

@inproceedings{grover1996fast,
  title={A Fast Quantum Mechanical Algorithm for Database Search},
  author={Grover, Lov K.},
  booktitle={Proceedings of the 28th Annual ACM Symposium on Theory of Computing},
  pages={212--219},
  year={1996}
}

@inproceedings{alghamdi2021future,
  title={The Future of Cryptocurrency Blockchains in the Quantum Era},
  author={Alghamdi, Sarah and Almuhammadi, Saleh},
  booktitle={IEEE International Conference on Blockchain},
  pages={544--551},
  year={2021},
  doi={10.1109/Blockchain53845.2021.00082}
}

@article{fernandez2020towards,
  title={Towards Post-Quantum Blockchain: A Review on Blockchain Cryptography Resistant to Quantum Computing Attacks},
  author={Fernández-Caramés, Tiago M. and Fraga-Lamas, Paula},
  journal={IEEE Access},
  volume={8},
  pages={21091--21116},
  year={2020},
  doi={10.1109/ACCESS.2020.2968986}
}

@article{elleithy2005denial,
  title={Denial of Service Attack Techniques: Analysis, Implementation and Comparison},
  author={Elleithy, Khaled M. and Blagovic, David and Wang, Cheng and Sideleau, Paul},
  journal={Journal of Systemics, Cybernetics and Informatics},
  volume={3},
  number={1},
  pages={66--71},
  year={2005}
}

@inproceedings{bhumichai2023feature,
  title={Feature Extraction of Network Traffic in Ethereum Blockchain Network Layer for Eclipse Attack Detection},
  author={Bhumichai, Dheerasak and Benton, Ryan},
  booktitle={SoutheastCon 2023},
  pages={869--876},
  year={2023},
  doi={10.1109/SoutheastCon51012.2023.10115126}
}

@article{diffie1976new,
  title={New Directions in Cryptography},
  author={Diffie, Whitfield and Hellman, Martin},
  journal={IEEE Transactions on Information Theory},
  volume={22},
  number={6},
  pages={644--654},
  year={1976},
  doi={10.1109/TIT.1976.1055638}
}

@article{nicolas2021blockchain,
  title={Blockchain System Defensive Overview for Double-Spend and Selfish Mining Attacks: A Systematic Approach},
  author={Nicolas, Kevin and Wang, Yufeng and Giakos, George C. and Wei, Bin and Shen, Haowei},
  journal={IEEE Access},
  volume={9},
  pages={3838--3857},
  year={2021},
  doi={10.1109/ACCESS.2020.3047365}
}

@inproceedings{anita2019blockchain,
  title={Blockchain Security Attack: A Brief Survey},
  author={Anita, N. and Vijayalakshmi, M.},
  booktitle={International Conference on Computing, Communication and Networking Technologies},
  pages={1--6},
  year={2019},
  doi={10.1109/ICCCNT45670.2019.8944615}
}

@incollection{douceur2002sybil,
  title={The Sybil Attack},
  author={Douceur, John R.},
  booktitle={Peer-to-Peer Systems},
  pages={251--260},
  year={2002},
  publisher={Springer}
}

@article{koblitz2015random,
  title={The Random Oracle Model: A Twenty-Year Retrospective},
  author={Koblitz, Neal and Menezes, Alfred J.},
  journal={Designs, Codes and Cryptography},
  volume={77},
  number={2},
  pages={587--610},
  year={2015}
}

@inproceedings{kaur2024hybrid,
  title={Hybrid Locking: An Effective Measure Against Reentrancy Attacks},
  author={Kaur, Navdeep and Kaur, Gurpreet and Kumar, Sanjeev and Saini, Harjinder Kaur},
  booktitle={Parul International Conference on Engineering and Technology},
  pages={1--6},
  year={2024},
  doi={10.1109/PICET60765.2024.10716087}
}

@article{alghamdi2024survey,
  title={A Survey of Blockchain Based Systems: Scalability Issues and Solutions, Applications and Future Challenges},
  author={Alghamdi, Turki A. and Khalid, Rabia and Javaid, Nadeem},
  journal={IEEE Access},
  volume={12},
  pages={79626--79651},
  year={2024},
  doi={10.1109/ACCESS.2024.3408866}
}

@inproceedings{wang2023defense,
  title={Defense against Sybil Attack in Blockchain Based on Improved Consensus Algorithm},
  author={Wang, Yifan and Tan, Min},
  booktitle={IEEE International Conference on Control, Electronics and Computer Technology},
  pages={986--989},
  year={2023},
  doi={10.1109/ICCECT57938.2023.10140278}
}

@inproceedings{alharby2017blockchain,
  title={Blockchain Based Smart Contracts: A Systematic Mapping Study},
  author={Alharby, Maher and Moorsel, Aad van},
  booktitle={International Conference on Artificial Intelligence and Soft Computing},
  pages={125--140},
  year={2017}
}

@article{bag2017bitcoin,
  title={Bitcoin Block Withholding Attack: Analysis and Mitigation},
  author={Bag, Soumya and Ruj, Sushmita and Sakurai, Kouichi},
  journal={IEEE Transactions on Information Forensics and Security},
  volume={12},
  number={8},
  pages={1967--1978},
  year={2017}
}

@article{saha2023protecting,
  title={Protecting the Decentralized Future: An Exploration of Common Blockchain Attacks and Their Countermeasures},
  author={Saha, Biprodip and Hasan, Md. Milon and Anjum, Nusrat and Tahora, Shaila and Siddika, Afsana and Shahriar, Hossain},
  journal={arXiv preprint arXiv:2306.11884},
  pages={1--29},
  year={2023},
  doi={10.48550/arxiv.2306.11884}
}

@article{cai2024assessing,
  title={Assessing and Neutralizing Multi-Tiered Security Threats in Blockchain Systems},
  author={Cai, Yize},
  journal={Applied and Computational Engineering},
  volume={49},
  pages={65--74},
  year={2024},
  doi={10.54254/2755-2721/49/20241063}
}

@article{das2024analysing,
  title={Analysing Attacks on Blockchain Systems in a Layer-based Approach},
  author={Das, Joydip and Tasin, Syed Ashraf Al and Rabbi, Md. Forhad and Ferdous, Md. Sadek},
  journal={arXiv preprint arXiv:2409.10109},
  year={2024}
}

@techreport{bai2020crystals,
  title={CRYSTALS-Dilithium Algorithm Specifications and Supporting Documentation},
  author={Bai, Shi and Ducas, Léo and Kiltz, Eike and others},
  institution={National Institute of Standards and Technology},
  year={2020},
  url={https://pq-crystals.org/dilithium/data/dilithium-specification-round3.pdf}
}

@misc{fouque2020falcon,
  title={Falcon: Fast-Fourier Lattice-Based Compact Signatures Over NTRU},
  author={Fouque, Pierre-Alain and Hoffstein, Jeffrey and Kirchner, Paul and others},
  year={2020},
  url={https://falcon-sign.info/falcon.pdf}
}

@inproceedings{bernstein2019sphincs,
  title={The SPHINCS+ Signature Framework},
  author={Bernstein, Daniel and Hülsing, Andreas and Kölbl, Stefan and others},
  booktitle={Proceedings of the ACM SIGSAC Conference on Computer and Communications Security},
  pages={2129--2146},
  year={2019},
  doi={10.1145/3319535.3363229}
}

@article{zhang2025pki,
  title={Research and Advances in PKI Technology},
  author={Zhang, Bin and Zhang, Yu and Zhang, Weizhe and others},
  journal={Journal of Software},
  pages={1--25},
  year={2025},
  doi={10.13328/j.cnki.jos.007305}
}

\end{document}